\documentclass[%
 reprint,
 amsmath,amssymb,
 aps,
]{revtex4-1}

\pdfoutput=1
\usepackage{graphicx}
\usepackage{dcolumn}
\usepackage{bm}
\usepackage{wrapfig}
\usepackage{color}
\usepackage[ruled,commentsnumbered]{algorithm2e}

\newcommand*{\bs}{\boldsymbol}

\newcommand*{\bl}{{\bs \ell}}
\newcommand{\sff}[1]{  \bs{\mathsf{#1}}  }

\begin{document}


\title{Shrinking the Quadratic Estimator}

\author{Ethan Anderes}
\thanks{Supported by NSF grant 1007480.}
\author{Debashis Paul}
\thanks{Supported by NSF grants DMR-1035468, DMS-1106690.}
\affiliation{%
Statistics Department\\ University of California, Davis, CA 95616}%




\begin{abstract}
We study a regression characterization for the quadratic estimator of weak lensing,  developed by Hu and Okamoto \cite{Hu2001b, HuOka2002, OkaHu2003},  for cosmic microwave background observations. 
 This characterization motivates a modification of the quadratic estimator by an adaptive Wiener filter which uses the robust Bayesian techniques described in \cite{bergerBook, bergerPaper, Straw}. This technique requires the user to propose a fiducial model for the spectral density of the unknown lensing potential but the resulting estimator is developed to be robust to misspecification of this model. 
 The role of the fiducial spectral density is to give the estimator superior statistical performance in a ``neighborhood of the fiducial model" while controlling the statistical errors when the fiducial spectral density is drastically wrong. 
 Our estimate also highlights some advantages provided by a Bayesian analysis of the quadratic estimator.
\end{abstract}

\maketitle


\section{Introduction}

The cosmic microwave background (CMB) measures temperature fluctuations in the early Universe some 400,000 years after the big bang. These fluctuations provide us with a picture of the Universe at the instant of recombination and contains a wealth of information for cosmology and cosmic structure. One important characteristic of the observed CMB is that the photon paths have been distorted, or lensed,  from the gravitational effect of intervening matter. 
Estimating this lensing is important for a number of reasons including, but not limited to, understanding cosmic structure, constraining cosmological parameters \cite{Kaplin, Smth2006} and detecting gravity waves \cite{knox2002, Kesden, SelH}.  
There have been a number of proposed estimators for the lensing of the CMB \cite{Hu2001b, HuOka2002, OkaHu2003,HiraSel2003a, HiraSel2003b,buncher, carv, anderes2011}. The most widely used  estimate  was developed in  \cite{Hu2001b, HuOka2002, OkaHu2003}  and is referred to as `the quadratic estimator'. This estimator, up to leading order, is an unbiased minimum variance estimator. In this paper we study the potential advantages obtained  by relaxing the unbiasedness constraint, borrowing statistical techniques from  Bayesian and regression theory.

 The effect of lensing is to simply remap the CMB, preserving surface brightness.   Up to leading order, the remapping  displacements are given by $\nabla \phi$, where $\phi$ denotes a lensing potential and is the planer projection of a three dimensional gravitational potential (see \cite{dod}, for example). The lensed CMB can then be written $\Theta(\bs x + \nabla \phi(\bs x))$ where $\Theta(\bs x)$ denotes the unlensed CMB temperature fluctuations projected to the observable sky. In this paper we work in the small angle limit  and use a flat sky approximation so that $\bs x\in \Bbb R^2$.   The isotropic and Gaussian assumptions for $\Theta(\bs x)$ translates to independence under the Fourier transform. However, for a fixed lensing potential $\phi$, the lensed CMB  becomes non-isotropic, leading to correlated Fourier modes. The quadratic estimator   takes advantage of this correlation and uses weighted sums of Fourier  cross products to unbiasedly (up to leading order) estimate the lensing potential.

  By relaxing the unbiasedness constraint  and utilizing robust Bayesian techniques (originally developed in \cite{bergerPaper,Straw} for statistical regression) we propose  an adaptive shrinkage adjustment to the nominal quadratic estimator.
The new estimator is given by 
\begin{equation}
\label{EQQ}
 \tilde \phi(\bs L) = \left[1- F_{L}^{\hat\phi} \frac{2 A_{\bs L}}{ C^{\phi\phi}_{L, \text{fid}}+ 2 A_{\bs L} } \right]\hat\phi(\bs L) 
 \end{equation}
where $\hat\phi(\bs L)$ is the regular quadratic estimator at frequency $\bs L\in\Bbb R^2$; $C^{\phi\phi}_{L, \text{fid}}$ is a fiducial spectral density model for the gravitational potential $\phi$; $2A_{\bs L}$ is the approximate variance of $\hat\phi(\bs L)$ derived in \cite{HuOka2002}; and $F_{L}^{\hat\phi}$ is an adaptive shrinkage factor defined in Section \ref{CCC}.
 The formula  makes it clear that the estimator is essentially an adaptive  Wiener filter: shrinking $\hat\phi(\bs L)$ when $2 A_{\bs L}/C^{\phi\phi}_{L, \text{fid}}$ is large and retaining $\hat\phi(\bs L)$ when $2 A_{\bs L}/C^{\phi\phi}_{L, \text{fid}}$ is small. 
The adaptive shrinkage factor $F_{L}^{\hat \phi}$ is derived as a posterior expected value in a robust Bayesian procedure and adapts the Wiener filter to account for uncertainty associated with misspecification of
 $C^{\phi\phi}_{L, \text{fid}}$. Indeed, the Bayesian viewpoint is the principal advantage of the estimate: using posterior draws of the lensing potential $\phi$, one can construct estimates and quantify the uncertainty for any non-linear function of the gravitational potential, including spectral density estimate. 
Moreover, in Section \ref{CCC} we show that the robust Bayesian procedure  is easy to simulate without resorting to expensive Markov Chain Monte Carlo (MCMC).

The non-informative prior for this Bayesian procedure is parameterized  by a fiducial spectral density  $C^{\phi\phi}_{L, \text{fid}}$ for the lensing potential $\phi$ and utilizes a hierarchal structure to induce robustness. Robustness, in this context,  pertains to the stability under mis-specification of the fiducial spectral density. We illustrate this property with simulations that demonstrate the similarity between $\tilde \phi(\bs L)$  and the optimal Wiener filter if one had access to  the true spectral density for $\phi$, even when the fiducial model is wrong by a relatively large factor  (similar robustness properties  have been demonstrated in \cite{bergerPaper}).   In fact, $\tilde \phi(\bs L)$ also has a frequentist interpretation that is independent of the Bayesian underpinnings. At the end of Section \ref{CCC} we discuss the connection  with a generalized James-Stein shrinkage estimator \cite{JS}, where the role of the fiducial spectral density is essentially to specify a shrinkage direction.

It should be noted that our Bayesian posterior is essentially an approximation
since  we model the lensing operation using the same first order Taylor approximation present in the quadratic estimator. Therefore, the well known bias in the quadratic estimator, derived in \cite{Kesden2, Han2010},  is still present in the estimate $\tilde \phi(\bs L)$ and in the associated posterior samples.   In the context of spectral density estimation one may be able to simply subtract this bias using the fiducial model $C^{\phi\phi}_{L, \text{fid}}$. However this would require some  stability of the bias under misspecification of $C^{\phi\phi}_{L, \text{fid}}$ which is yet to be established. 

The remainder of this paper is organized as follows. In Section \ref{Reg} we establish the connection between regression and the quadratic estimator. This connection is then used in Section \ref{SQE1} to derive the optimal Wiener filter of the quadratic estimate when the true spectral density for $\phi$ is known. In Section \ref{SQE2} we  derive the adaptive shrinkage estimator (\ref{EQQ}) using  Bayesian techniques when the spectral density for $\phi$ is unknown. Finally, in Section \ref{Sim} we present some simulation examples  exploring the advantages of the estimator (\ref{EQQ}) and its Bayesian interpretation.

\section{Regression and the quadratic estimator}
\label{Reg}
In this section we characterize the quadratic estimator in terms of generalized least squares regression. The main utility of this connection is the incorporation of tools from statistical regression that lead to the shrinkage estimators given in sections \ref{SQE1} and \ref{SQE2}. 

The quadratic estimator is derived under the assumption that the observed lensed CMB field is contaminated by additive noise and an instrumental beam. Throughout this paper we let  $\tilde \Theta(\bs x)$ denote the observed CMB field  with lensing, beam (denoted by $\varphi$) and noise (denoted $N(\bs x)$) so that  
\[ \tilde \Theta(\bs x)= N(\bs x)+\int \Theta(\bs y+\nabla \phi(\bs y)) \varphi(\bs x - \bs y) \, d\bs y. \]
The quadratic estimator is based on a first order Taylor approximation in $\nabla \phi $ on the lensed CMB field:
$\Theta(\bs x+\nabla \phi(\bs x))= \Theta(\bs x) + \nabla \phi(\bs x)\cdot \nabla \Theta(\bs x)+ O(\phi^2)$. 
By truncating $O(\phi^2)$ terms one gets the following approximation for the correlation of Fourier modes induced by lensing:
\begin{align}
\label{FundEQ}
\Bigl\langle \tilde \Theta(\bs \ell +\bs L) \tilde \Theta(\bs \ell )^*  \Bigr\rangle  \approx \delta_{\bs L} C^{\Theta\Theta}_{\bl, \text{expt}} + f_{\bs L}(\bl) \phi(\bs L)
\end{align}
 where  $f_{\bs L}(\bl) \equiv (2\pi)^{-1}\bigl[  ( \bl+\bs L)\cdot \bs L\,  C^{\Theta\Theta}_{\bl+\bs L}  - \bl \cdot \bs L \, C^{\Theta\Theta}_{\bl} \bigr]\varphi(\bl + \bs L) \varphi^*(\bl) $,  $C^{\Theta\Theta}_{\bs \ell}$ denotes the flat sky power spectrum for $\Theta$ and 
  $C^{\Theta\Theta}_{\bl,\text{expt}}\equiv |\varphi(\bl)|^2 C^{\Theta\Theta}_\bl+C^{NN}_\bl$ so that $C^{\Theta\Theta}_{\bl, \text{expt}}$ denotes the power spectrum  for the unlensed CMB corrupted with experimental noise and beam. To avoid any confusion notice that we are adopting the notational convention of \cite{LewisReview} (see Section 4.1) so that $C^{\Theta\Theta}_{\bs \ell}$ is defined by $\langle \Theta(\bl)\Theta(\bl^\prime)^* \rangle =\delta_{\bl-\bl^\prime} C^{\Theta\Theta}_{\bl} $ where $\Theta(\bl)\equiv \int  e^{-i\bs x\cdot \bl} \Theta(\bs x) \frac{d^2\bs x}{2\pi}$ and  $\delta_\bl \equiv \int  e^{i\bs x\cdot \bl} \frac{d^2\bs x}{(2\pi)^2}$. 
 Throughout the remainder of the paper we let angled brackets (as used in (\ref{FundEQ})) denote expectation over the unlensed CMB and the instrumental noise, while holding the potential $\phi$ fixed.


Equation (\ref{FundEQ}) exposes the nonstationarity in the lensed temperature field through the cross correlation of Fourier modes.
The quadratic estimator uses this correlation to estimate $\phi(\bs L)$ by weighted averaging cross products of $\tilde \Theta(\bl)$ separated at lag $\bs L\neq 0$. In particular, let $k=1, 2, \ldots$ index a set of frequencies $\bs \ell_k\in\Bbb R^2$.  Now, for each $k$ define the normalized cross product at lag $\bs L$, $ \mathsf v_{\bs L,k}\equiv { \tilde\Theta (\bl_k+\bs L) \tilde \Theta(\bl_k)^*}/{f_{\bs L}(\bl_k)}$  and  the corresponding vector of these cross products  $\bs{\mathsf{ v}}_{\bs L}\equiv  (\mathsf{ v}_{\bs L,1}, \mathsf{ v}_{\bs L,2}, \ldots)^\dagger$. Since $\bs L \neq 0$, equation (\ref{FundEQ}) implies that each $\mathsf v_{\bs L, k}$ is a noisy unbiased estimate of $\phi(\bs L)$  (up to the approximation in (\ref{FundEQ})). Writing this statement in a regression setting one obtains
\begin{align}
\label{dataEQ}
 \sff v_{\bs L}&=  \bs 1\phi(\bs L) + \sff e_{\bs L}
  \end{align}
 where for each fixed $\bs L$, $ \sff e_{\bs L} $ is an error vector and $\bs 1$ is a vector of ones.
Approximation (\ref{FundEQ}) establishes that $\langle\sff e_{\bs L}\rangle\approx 0$. The generalized least squares regression estimator for $\phi(\bs L)$ is then
\begin{equation}
\label{Prelim} 
\hat \phi(\bs L)\equiv  (\bs 1^\dagger\bs{\mathsf{N}}^{-1}_{\bs L}\bs 1)^{-1} \bs 1^\dagger\bs{\mathsf{N}}_{\bs L}^{-1} \sff v_{\bs L}
\end{equation}
where $\bs{\mathsf N}_{\bs L}$ is the covariance matrix of $\sff e_{\bs L}$  which is approximated as follows:
\begin{align*}
 \bigl(\sff N_{\bs L}\bigr)_{k,j} &\equiv\langle \mathsf v^{\phantom{*}}_{\bs L, k} \mathsf v_{\bs L, j}^* \rangle - |\phi(\bs L)|^2\\
 & \approx [\delta^2_{\bl_k-\bl_j}+ \delta^2_{\bl_k+\bl_j+\bs L}] \frac{C^{\Theta \Theta}_{\bl_k+\bs L,\text{expt}}C^{\Theta\Theta}_{\bl_k,\text{expt}}}{f_{\bs L}(\bl_k)f_{\bs L}(\bl_j)^* }.
\end{align*}
 The above approximation is obtained from (\ref{FundEQ}) using Wick's theorem and discarding any $O(\phi)$ terms. Also notice that, in practice, the covariance matrix $\sff N_{\bs L}$ is based on a discrete grid approximation in Fourier space that arises from finite sky observations of $\tilde\Theta$.  In particular, $\delta_\bl$ is approximated as $1/ \Delta \bs L$ when $\bl =0$ and zero otherwise, where $\Delta \bs L$ is the area element of the grid in Fourier space. For the remainder of the paper we do not distinguish the discrete case versus the continuous and simply equate $\delta_0$ with $1/ \Delta \bs L$ leaving it understood that equality holds in the limit as $\Delta \bs L\rightarrow 0$.

This is not the typical derivation of the quadratic estimator. However, it should be no surprise  they are related since both are minimum variance estimators. To connect the two, notice that the only reason $\delta_{\bl_k+\bl_j+\bs L}$ appears in $\sff N_{\bs L}$ is that the terms $\tilde\Theta(\bl+\bs L)  \tilde\Theta(\bl)^*/f_{\bl}(\bs L)$ and $\tilde \Theta(\bl^\prime+\bs L)  \tilde \Theta(\bl^\prime)^*/f_{\bl^\prime}(\bs L)$ are the same when $\bl +\bl^\prime + \bs L=0$. Therefore, if we only include unique observations in $\sff v_{\bs L}$, $\sff N_{\bs L}$ becomes diagonal.
It is now easy to see that
 \begin{align}
 \label{asdf1}
& {  \bs 1^\dagger\sff N_{\bs L}^{-1} \sff v_{\bs L}}  =\frac{1}{2\delta_0} \int  \frac{f_{\bs L}(\bl)^* { \tilde \Theta(\bs \ell+\bs L)  \tilde\Theta (\bs \ell)^*}}{C^{\Theta\Theta}_{\bl+\bs L,\text{expt}}C^{\Theta\Theta}_{\bl,\text{expt}}} \,d\bl   \\
\label{asdf2}
&{ \bs 1^\dagger\sff N_{\bs L}^{-1}\bs 1} = \frac{1}{2\delta_0}\int   \frac{|f_{\bs L}(\bl)|^2}{C^{\Theta\Theta}_{\bl+\bs L,\text{expt}}C^{\Theta\Theta}_{\bl,\text{expt}}}\,d\bl. 
\end{align}
where the equalities are understood to hold in the continuous-to-discrete approximation inherent in $\delta_0$ and the above integrals.
 Notice that the factor of $1/2$ comes from the fact that the integrals have non-unique terms in the integrand.
Returning to the original characterization of the quadratic estimator developed in \cite{HuOka2002,Hu2001b}, $\hat \phi$ has the form $A_{\bs L} \int_{\Bbb R^d}    \tilde \Theta(\bs \ell+\bs L)  \tilde \Theta(\bs \ell)^*g_{\bl}(\bs L)\, {d\bl} $ where the weights $g$ must satisfy the constraint $g_{\bs L}(\bl) f_{\bs L}(\bl)\geq 0$ and the normalizing constant $A_{\bs L}$ is given by $A_{\bs L}^{-1}\equiv\int {g_{\bs L}(\bl) f_{\bs L}(\bl)}  \,d\bl $ to ensure unbiasedness. In \cite{Hu2001b} the optimal weights $g$ are derived to  be proportional to ${f_{\bs L}(\bl)^*}/[{C^{\Theta\Theta}_{\bl+\bs L,\text{expt}}C^{\Theta\Theta}_{\bl,\text{expt}}}] $. Now, plugging in these optimal weights and using (\ref{asdf1}) and (\ref{asdf2}) one immediately gets
 \begin{align*}
  \hat\phi(\bs L)  &\equiv  (\bs 1^\dagger \sff N_{\bs L}^{-1}\bs 1)^{-1} \bs 1^\dagger \sff N_{\bs L}^{-1} \bs{\mathsf v}_{\bs L}\\
  &=A_{\bs L} \int_{\Bbb R^d}    \tilde \Theta(\bs \ell+\bs L)  \tilde \Theta(\bs \ell)^*g_{\bs L}(\bl)\, {d\bl} .
\end{align*}
Moreover, the  prediction variance  follows easily from
standard regression theory which gives the following variance of  $\hat \phi$
\begin{align*} \bigl\langle \hat \phi(\bs L) \hat \phi(\bs L^\prime)^* \bigr\rangle - & \phi(\bs L)\phi(\bs L^\prime)^*\\
&=\begin{cases}
  (\bs 1^\dagger\bs{\mathsf{N}}^{-1}_{\bs L}\bs 1)^{-1}, &  \text{if $\bs L=\bs L^\prime$;} \\
  0, &\text{otherwise}
 \end{cases} \\
  &=  \delta_{\bs L-\bs L^\prime}{2}{A_{\bs L}}
 \end{align*}
where the last equality holds by (\ref{asdf2}).
 
\section{Shrinking the quadratic estimator when $C^{\phi\phi}_L$ is known}
\label{SQE1}

The main reason the regression characterization of the quadratic estimator is useful is to easily see how one would relax the unbiasedness constraint when the true $C^{\phi\phi}_L$ is known. Indeed, a natural extension to the generalized regression estimate is called ridge regression which can be written
\begin{equation}
\label{ridge}  
 \tilde\phi_{\lambda_{\bs L}}(\bs L)\equiv (\bs 1^\dagger\sff N_{\bs L}^{-1}\bs 1 + \lambda_{\bs L}^{-1} )^{-1} \bs 1^\dagger\sff N_{\bs L}^{-1}\sff v_{\bs L}
 \end{equation}
where  $\lambda_{\bs L}$ is a ridge parameter which regularizes the intrinsic instability that arrises when $\bs 1^\dagger\sff N_{\bs L}^{-1}\bs 1$ is small (i.e. when the variance of the re-construction is large).
When $\phi(\bs L)$ is a realization from a mean zero Gaussian random field and $\lambda_{\bs L}$ is set to the variance of $\phi(\bs L)$ then the ridge regression estimate $ \tilde\phi_{\lambda_{\bs L}}(\bs L)$ can be interpreted as the posterior expected value given the data $\sff v_{\bs L}$ under the assumption that both $\sff e_{\bs L}$ (defined in (\ref{dataEQ})) and $\phi(\bs L)$ are uncorrelated, mean zero and jointly Gaussian. 
Returning to the original characterization of the quadratic estimate this leads to  the following expression for $\tilde\phi_{\lambda_{\bs L}}(\bs L) $:
 \begin{align*}
 \tilde\phi_{\lambda_{\bs L}}(\bs L) 
=  \frac{1}{A_{\bs L}^{-1}+  2/C_{L}^{\phi\phi} }  \int_{\Bbb R^d}     \tilde \Theta(\bs \ell+\bs L)  \tilde \Theta(\bs \ell)^* g_{\bs L}(\bl) &\,{d\bl} 
\\
\text{when setting $\lambda_{\bs L}=\delta_0 C^{\phi\phi}_L$.}&\nonumber
 \end{align*}
Moreover,  the posterior variance of $\phi(\bs L)$, under this Bayesian interpretation, is simply $(\bs 1^\dagger\sff N_{\bs L}^{-1}\bs 1 + \lambda_{\bs L}^{-1} )^{-1} =  \delta_0/\bigl[ (2 A_{\bs L})^{-1} + (C^{\phi\phi}_L)^{-1} \bigr]$.

The above Bayesian interpretation of $\tilde\phi_{\lambda_{\bs L}}(\bs L) $ relies on  the assumption that $\mathsf v_{\bs L,k}$ is Gaussian.  This is clearly violated since $\mathsf v_{\bs L,k}$ is a product of two Gaussians. However,  one can also interpret the ridge regression estimate $\tilde\phi_{\lambda_{\bs L}}(\bs L)$ in the spirit of a Wiener filter of the quadratic  estimator which makes the Gaussianity assumption less worrisome. In particular, if one treats the quadratic estimate $\hat\phi$ as data, the approximate unbiasedness of the quadratic estimator allows one to write
\begin{align} 
\label{data}
&\hat\phi(\bs L)=\phi(\bs L) + \epsilon(\bs L) 
\end{align}
where $\langle\epsilon(\bs L)\rangle \approx 0$ and $\bigl\langle\epsilon(\bs L)\epsilon(\bs L^\prime)^*\bigr\rangle =\delta_{\bs L - \bs L^\prime}2A_{\bs L}$. 
The supposition that $\epsilon$ is Gaussian becomes more believable since $\hat \phi$ is a weighted average of $\mathsf v_{\bs L,k}$ which are  approximately independent random variables (up to zero order in $\phi$). If one now applies the Bayesian paradigm under the assumption that $\phi$ and $\epsilon$ are uncorrelated mean zero Gaussian random fields, then the posterior expected value of $\phi$ when observing $\hat\phi$ becomes
\begin{align} 
\label{weiner}
\phantom{XXXXX}\tilde \phi_{\lambda_{\bs L}}(\bs L) =\frac{C_{L}^{\phi\phi}}{2A_{\bs L}+  C_{L}^{\phi\phi} }  \hat \phi(\bs L)\phantom{XXXXXX}& 
\\
\text{when setting $\lambda_{\bs L}=\delta_0 C^{\phi\phi}_L$.}&\nonumber
\end{align}
This agrees with (\ref{ridge}) and is clearly recognized as a Wiener filter of the quadratic estimate $\hat \phi$.

\section{Adaptive shrinking when  $C^{\phi\phi}_L$ is unknown}
\label{SQE2}

The previous section derived the optimal Wiener filter of the quadratic estimate when the spectral density $C^{\phi\phi}_L$ is known.  
In this section we study the scenario that $C^{\phi\phi}_L$ is not known.
The main question is how to derive a shrinkage factor, as in the Wiener filter (\ref{weiner}), without knowledge of $C^{\phi\phi}_{L}$. This is derived using  a hierarchal Bayesian analysis which establishes that the shrinkage factor is a mixture over the possible values of the $C^{\phi\phi}_L$ supported by the data. In   \ref{TTT}  we first discuss the  full Bayesian analysis  which requires a prior distribution on the unknown spectral density $C^{\phi\phi}_L$. To circumvent computational difficulties with such an analysis we then recommend a non-informative generalized prior, in \ref{CCC}, which gives rise our robust adaptive shrinkage estimator (\ref{EQQ}) presented in the introduction. 

 To align our notation with standard statistical theory   it will be useful to concatenate the values of $\hat \phi(\bs L)$, for different frequencies $\bs L$,
into one  data vector of length $n$, denoted $\sff{\hat \phi}$. Define $\sff \phi$ similarly as the vector of  $\phi(\bs L)$ values for the matching frequencies $\bs L$ used to construct $\sff{\hat \phi}$. 
We make the additional assumption that $\sff \phi$ and $\sff{\hat \phi}$  contain only unique elements up to complex conjugation (so there are no distinct coordinates with values $w$ and $z$ such that $z=w$ or $z=w^*$).
 Working with the vectors $\sff \phi$ and $\sff{\hat \phi}$, instead of the functions $\phi$ and $\hat\phi$, has the additional  advantage that one can easily extend to the case where the adaptive shrinkage is done on separate annuli in Fourier space (which is discussed in Remark 1 at the end of Section \ref{CCC}).  

We work under the viewpoint, used to derived equation (\ref{weiner}), that treats $\sff{\hat \phi}$ as  ``data"  which are then used to estimate $\sff \phi$. Translating equation (\ref{weiner}) to our vector notation, the Weiner filter  is given simply by  matrix multiplication
 \begin{equation}
 \label{wnr}
  \sff{\tilde\phi}_\Lambda \equiv \Lambda(\Lambda + \Sigma)^{-1}\sff{\hat\phi}
  \end{equation}
  where   the matrices $\Sigma$ and $\Lambda$ are defined by
\begin{align*}
\Sigma &\equiv \Bigl(\delta_{\bs L_k- \bs L_j} 2A_{\bs L_k}\Bigr)_{k,j=1}^n\\
\Lambda &\equiv \Bigl(\delta_{\bs L_k- \bs L_j} C^{\phi\phi}_{L_k}\Bigr)_{k,j=1}^n.
\end{align*}
In addition, our notation allows us to clearly write the relationship between $\sff \phi$ and $\sff{\hat\phi}$ in a hierarchal Bayesian setting
\begin{align}
\label{f1}
\sff{\hat\phi} \bigl| &\sff \phi \sim \mathcal N\left(\sff \phi, \Sigma \right )  \\
\sff{\phi} \bigl|&\Lambda \sim \mathcal N\left(0,  \Lambda  \right)  \label{ff1}
\end{align}
where $\sff \phi \bigl|\Lambda \sim \mathcal N(0, \Lambda)$ means $\text{\small Re}(\sff{\phi})$ and $\text{\small Im}(\sff{\phi})$ are independent Gaussian random vectors with individual distributions given by $\text{\small Re}(\sff{\phi})\sim \mathcal N\left(0,  \Lambda/2\right) $ and $\text{\small Im}(\sff{\phi})\sim \mathcal N\left(0,  \Lambda/2\right)$. 

\subsection{The Bayes Solution}
\label{TTT}

The  Bayesian paradigm is the clearest way to understand how one adapts  (\ref{wnr}) when $\Lambda$ is unknown. Indeed, if one is willing to model the  uncertainty in $\Lambda$ (equivalently in $C^{\phi\phi}_L$) using a prior probability density $\mathcal P(\Lambda)$, then Bayes theorem in conjunction with (\ref{f1}) and (\ref{ff1}) gives a posterior density  $\mathcal P(\sff \phi,\Lambda |\hat{\sff \phi} )\propto \mathcal P (\hat{\sff \phi}|\sff \phi  )\mathcal P(\sff \phi |\Lambda)\mathcal P(\Lambda)$. 
The posterior density $P(\sff \phi,\Lambda |\hat{\sff \phi} )$ quantifies the joint uncertainty in the unobserved $\sff \phi$ and $\Lambda$ when observing the data $\sff{\hat\phi}$.
Notice that one can marginalize out $\Lambda$ to obtain a posterior on $\sff{\phi}$ exclusively, 
  $\mathcal P(\sff \phi |\hat{\sff \phi} )= \int \mathcal P(\sff \phi,\Lambda |\hat{\sff \phi} )d\Lambda =\int \mathcal P(\sff \phi |\Lambda,\hat{\sff \phi} ) \mathcal P(\Lambda|\hat{\sff \phi}) d\Lambda $ where $d\Lambda$ corresponds to coordinate-wise area element. Now the posterior expected value of $\sff \phi$ given the data $\sff{\hat \phi}$ can be computed as 
\begin{align}
\nonumber
\int \sff \phi   \mathcal P(\sff \phi |{\sff {\hat\phi} }) \,d\sff \phi&= 
\int \underbrace{\Bigl[\int  \sff \phi   \mathcal P(\sff \phi |\Lambda,{\sff {\hat\phi} })\,d\sff\phi\Bigr]}_\text{\small Weiner filter (\ref{wnr})}   \mathcal P(\Lambda | {\sff {\hat\phi} })\,d \Lambda \\
&=  \int \sff{\tilde\phi}_\Lambda    \mathcal P(\Lambda | {\sff {\hat\phi} })\,d\Lambda.  \label{eeee}
\end{align}
The advantage of (\ref{eeee}) is that it clearly exposes how to handle  the situation when $\Lambda$ is unknown: average  the Wiener filter $\sff{\tilde\phi}_\Lambda$ over different possibilities for $\Lambda$ supported by the data (through the posterior $\mathcal P (\Lambda|\sff{\hat \phi})$).
 In fact, since $\sff{\tilde\phi}_\Lambda$ depends on $\Lambda$ only through a multiplicative factor, (\ref{eeee}) simplifies to 
\begin{equation} 
\int \sff{\tilde\phi}_\Lambda    \mathcal P(\Lambda | \sff{\hat\phi} )\,d\Lambda = \underbrace{ \left[\int \Lambda(\Lambda + \Sigma)^{-1} \mathcal P(\Lambda|\sff{\hat\phi}) \, d\Lambda. \right]}_{\substack{\text{\small posterior expected} \\ \text{\small shrinkage factor} }} \sff{\hat\phi} \label{fffran}
\end{equation}
Therefore, to account for uncertainty in $\Lambda$ when estimating $\sff \phi$ simply replace the shrinkage factor $\Lambda (\Lambda+\Sigma)^{-1}$ in (\ref{wnr}) with it's expected value under the posterior distribution on $\Lambda$, $\mathcal P(\Lambda|\sff{\hat\phi})$.

\subsection{Robust generalized Bayesian adaptive shrinkage} \label{CCC}

The Bayesian analysis, while being a complete probabilistic combination of the data $\sff{\hat \phi}$ and the prior knowledge for $\Lambda$, can be time consuming for two reasons. First, one needs to translate the knowledge in $\Lambda$ to a prior distribution $\mathcal P(\Lambda)$. Secondly, getting posterior samples from $\mathcal P(\Lambda|\sff{\hat\phi})$ often requires advanced Monte Carlo techniques (similar to those found in \cite{eriksen, Wandelt} for example). For these reasons we present a simplified non-informative generalized prior on $\Lambda$, originally developed in \cite{bergerPaper, Straw}, for quick exploratory analysis.  This prior requires the user to specify a  fiducial spectral density model, denoted $C^{\phi\phi}_{L,\text{fid}}$,  and is designed to be both robust against alternative spectral truths and computationally simple.

To specify the prior start by defining the following matrix based on the fiducial spectral density 
\begin{align*}
\Lambda_\text{fid} &\equiv \Bigl(\delta_{\bs L_k- \bs L_j} C^{\phi\phi}_{L_k, \text{fid}}\Bigr)_{k,j=1}^n.
\end{align*}
The non-informative generalized prior for $\Lambda$ is given by 
\begin{align}
&\Lambda \sim  \xi (\Sigma+ \Lambda_\text{fid})- {\Sigma} \label{RB1}\\
&\xi \text{ has generalized density $\propto \xi^{-1}$ on $(\rho,\infty)$} \label{RB2} 
\end{align}
where $\rho$ is the largest eigenvalue of $\Sigma (\Sigma+ \Lambda_\text{fid})^{-1}$ (which ensures that $\xi (\Sigma+ \Lambda_\text{fid})- {\Sigma} $ is always positive definite). 
Since the support of $\xi$ contains $1$, the prior includes the fiducial model $\Lambda_\text{fid}$ as a possible truth. The generalized density $\xi^{-1}$ is an improper prior (it integrates to infinity) which can be derived as the well known Jeffereys' non-informative prior within the class of distributions $\Lambda \sim \xi (\Sigma+ \Lambda_\text{fid})- {\Sigma}$ for the hierarchal parameter $\xi$.
The theoretical properties of this and similar priors have been extensively studied in the statistical literature (see \cite{bergerBook, bergerPaper, Straw}) and has been shown to yield estimators with desirable statistical properties.

One of the advantages of this prior is that marginalizing over $\sff{\phi}$ and $\Lambda$ collapses (\ref{f1}), (\ref{ff1}) and (\ref{RB1})  to
\begin{align}
&\sff{\hat\phi}\bigl|\xi \sim \mathcal N\bigl(0,\xi (\Sigma+ \Lambda_\text{fid}) \bigr). \label{RB3}  
\end{align}
Now it is easy to obtain the posterior distribution on $\xi$. In particular, let $\mathcal P(\sff{\hat\phi}|\xi, \Lambda_\text{fid}, \Sigma) $ denote the marginal density of $\sff{\hat\phi}$ in (\ref{RB3}) so that
\begin{align*}
 &\mathcal P(\sff{\hat\phi}|\xi, \Lambda_\text{fid}, \Sigma) \propto \frac{1}{\xi^{n}|\Sigma +\Lambda_\text{fid}|} \exp\left (-\|\sff{\hat\phi}  \|^2_\text{fid}/{\xi} \right)
 \end{align*}
 where $\|\sff{\hat\phi}  \|^2_\text{fid}$ is defined by
 \begin{align*}
 \|\sff{\hat\phi}  \|^2_\text{fid}\equiv &\left\| (\Sigma +\Lambda_\text{fid})^{-1/2} \text{\small Re}(\sff{\hat \phi})\right\|^2\\
 &\quad+  \left\|(\Sigma +\Lambda_\text{fid})^{-1/2} \text{\small Im}(\sff{\hat \phi})\right\|^2.
 \end{align*}
  Note,
the number of elements in $\sff{\hat\phi}$ is $n$  (so that the vector $(\text{\small Re}(\sff{\hat\phi})^\dagger, \text{\small Im}(\sff{\hat\phi})^\dagger)^\dagger$ has $2n$ entries, for example).
Therefore a formal application of Bayes theorem for $\xi$ under the model (\ref{RB3}) gives
\begin{align}
\mathcal P\bigl(\xi|\sff{\hat\phi}\bigr) &\propto    \mathcal P(\sff{\hat\phi}|\xi, \Lambda_\text{fid}, \Sigma)\mathcal P\bigl(\xi) \nonumber \\
& \propto \frac{1}{\xi^{n+1}}  \exp\left (-\|\sff{\hat\phi}  \|^2_\text{fid}/{\xi} \right)   \label{nnorm}
\end{align}
on $(\rho,\infty)$ where $\mathcal P(\xi)\propto \xi^{-1}$ denotes the prior for $\xi$.
This shows that the posterior $\mathcal P(\xi|\sff{\hat\phi})$ has a truncated inverse gamma distribution, which can easily be sampled from, as in Algorithm \ref{alg1} below.

\LinesNumbered
\begin{algorithm}
\DontPrintSemicolon
 \caption{Sample from  $\mathcal P(\xi|\sff{\hat\phi})$\label{alg1}}
  :\hspace{.2cm}{Simulate a Gamma random variable $\zeta$ with density proportional to ${\zeta^{n-1}}\exp(-\zeta \|\sff{\hat\phi}  \|^2_\text{fid})$ \label{step1}}\;
 :\hspace{.2cm} {\textbf{if} ${\zeta^{-1}} (\Sigma+ \Lambda_\text{fid})- {\Sigma}$ is positive definite \textbf{then} go to step 3, \textbf{else} go back to step 1}\;
 :\hspace{.2cm}{\textbf{return} $\xi\leftarrow\zeta^{-1}$}
\end{algorithm}


The posterior samples from $\mathcal P(\xi|\sff{\hat\phi})$  do not have a direct physical interpretation. However,  Algorithm \ref{alg2} shows how sampling from $\mathcal P(\xi|\sff{\hat\phi})$ allows easy sampling from $\mathcal P(\sff \phi | \sff{\hat\phi})$.  Actually, it is not immediately  obvious that Algorithm 2 gives samples from $\mathcal P(\sff \phi|\sff{\hat\phi})$ since the prior $\mathcal P(\xi)$ is improper. However, a careful application of Fubini shows, indeed,  Algorithm 2  samples $\mathcal P(\sff \phi|\sff{\hat\phi})$.
In addition,  the posterior expected shrinkage factor in  (\ref{fffran})  can be computed as follows
\begin{align}
\int \Lambda(\Sigma+\Lambda)^{-1}&\mathcal P({\Lambda|\sff{\hat\phi}}) \nonumber\\
& =\int \left[I-\xi^{-1} \Sigma(\Sigma+\Lambda_\text{fid})^{-1}  \right] \mathcal P(\xi|\sff{\hat\phi})d\xi\nonumber\\
&=I-F^{\sff{\hat\phi}}\, \Sigma(\Sigma+\Lambda_\text{fid})^{-1}  \label{rrr}
 \end{align}
where $ F^{\sff{\hat\phi}} \equiv \int \xi^{-1}  \mathcal P({\xi|\sff{\hat\phi}}) $ and $I$ denotes the $n\times n$ identity matrix.
Keeping track of the normalization factor  in (\ref{nnorm}), one obtains the following analytic expression for $F^{\sff{\hat\phi}}$
\begin{equation} 
\label{venn}
F^{\sff{\hat\phi}} =   \frac{n-1}{ \| \sff{\hat\phi}\|^2_\text{fid}} \frac{P\bigl(n,  \| \sff{\hat\phi}\|^2_\text{fid}/\rho\bigr)}{P\bigl(n-1,  \| \sff{\hat\phi}\|^2_\text{fid}/\rho\bigr)} \end{equation}
where
 $P(a,x)$ is the normalized incomplete gamma function given by   $P(a,x)\equiv  \frac{1}{\Gamma(a)}  \int_{0}^x t^{a-1}e^{-t} dt $. Combining equations (\ref{rrr}) and (\ref{fffran}) one obtains the following adaptive shrinkage estimate of $\sff \phi$ given $\sff{\hat\phi}$
 \begin{align}
 \label{bff}
\sff{ \tilde \phi} &\equiv \int \sff \phi \mathcal P(\sff \phi |\sff{\hat\phi}) = \bigl[ I-F^{\sff{\hat\phi}}\, \Sigma(\Sigma+\Lambda_\text{fid})^{-1} \bigr] \sff{\hat\phi}
   \end{align}
   which is recognized as the matrix form of (\ref{EQQ}).

{\em Remark 1:} In our implementation of  (\ref{bff}) we partition the Fourier frequencies $\bs L$ into concentric annuli around the origin and construct the posteriors $\mathcal P(\xi | \sff{\hat\phi})$  and $\mathcal P(\sff \phi | \sff{\hat\phi})$ separately on each annuli.  In this way we obtain distinct shrinkage factor adjustments $F^{\sff{\hat\phi}}$ for each annuli, hence the dependence of $F^{\sff{\hat\phi}}$ on $L$, written $F_{L}^{\sff{\hat\phi}}$ in (\ref{EQQ}). This essentially adds flexibility by allowing independent priors  $\mathcal P(\xi)$ for each annuli. In fact, if $2A_{\bs L}$ changes drastically within an annuli, one may  further partition with the goal of obtaining partitions within which the values of $2A_{\bs L}$ are similar.

{\em Remark 2:} Although, one has access to an analytic expression for the $F^{\hat\phi}$ given in  (\ref{venn}), this formula becomes numerically unstable when either $\|\sff{\hat\phi}  \|^2_\text{fid}$ is small or $n$ large. Therefore we recommend approximating $F^{\hat\phi}_L=\int \xi^{-1}\mathcal P(\xi |\sff{\hat\phi}) d\xi$  by averaging samples of $\xi^{-1}$ obtained from Algorithm \ref{alg1}.

\LinesNumbered
\begin{algorithm}
\DontPrintSemicolon
 \caption{Sample from  $\mathcal P(\sff \phi|\sff{\hat\phi})$\label{alg2} }
  :\hspace{.2cm}{Simulate $\xi$ from Algorithm \ref{alg1}}\;
 :\hspace{.1cm} {Set $\Lambda \leftarrow {\xi} (\Sigma+ \Lambda_\text{fid})- {\Sigma}$ and simulate \[\sff \phi\sim \mathcal N\Bigl(\Lambda(\Sigma+\Lambda)^{-1}\sff{\hat\phi}, \, [\Sigma^{-1} + \Lambda^{-1}]^{-1}\Bigr)\]}\;
 :\hspace{.2cm}{\textbf{return} $\sff \phi$}
\end{algorithm}

%

\begin{figure*}[t]
\includegraphics[height = 1.7in]{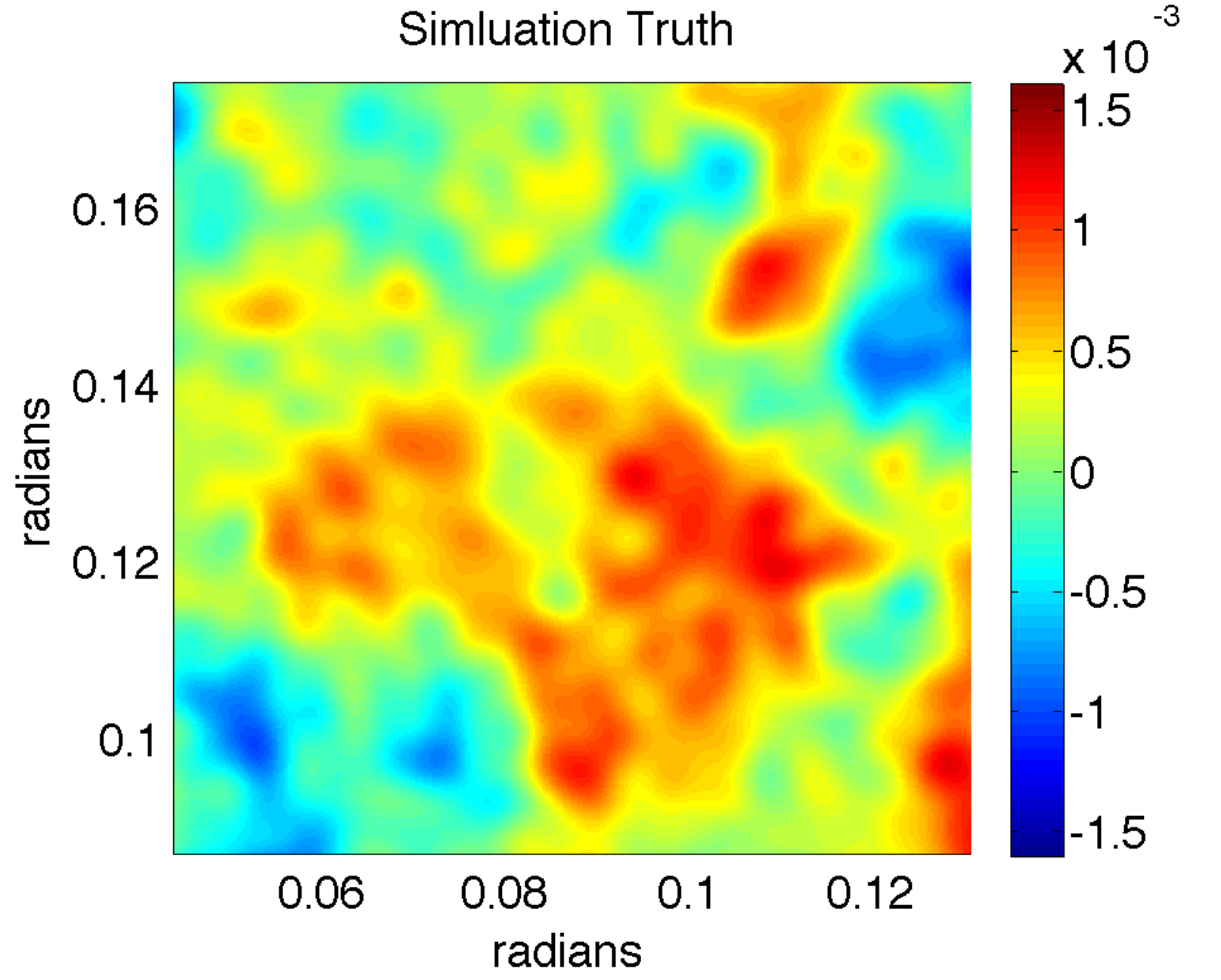}\\
\includegraphics[height = 1.7in]{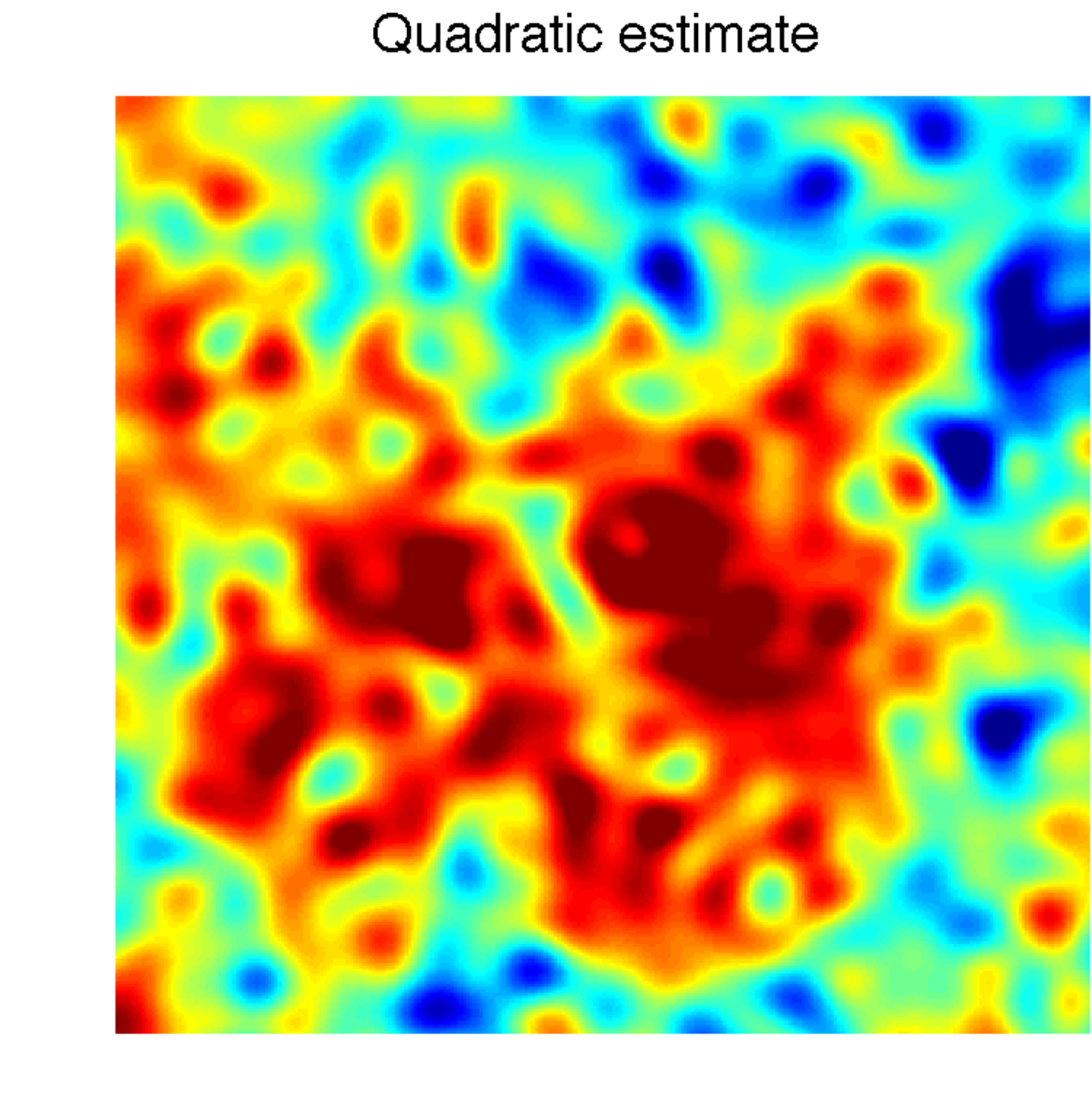}
\includegraphics[height = 1.7in]{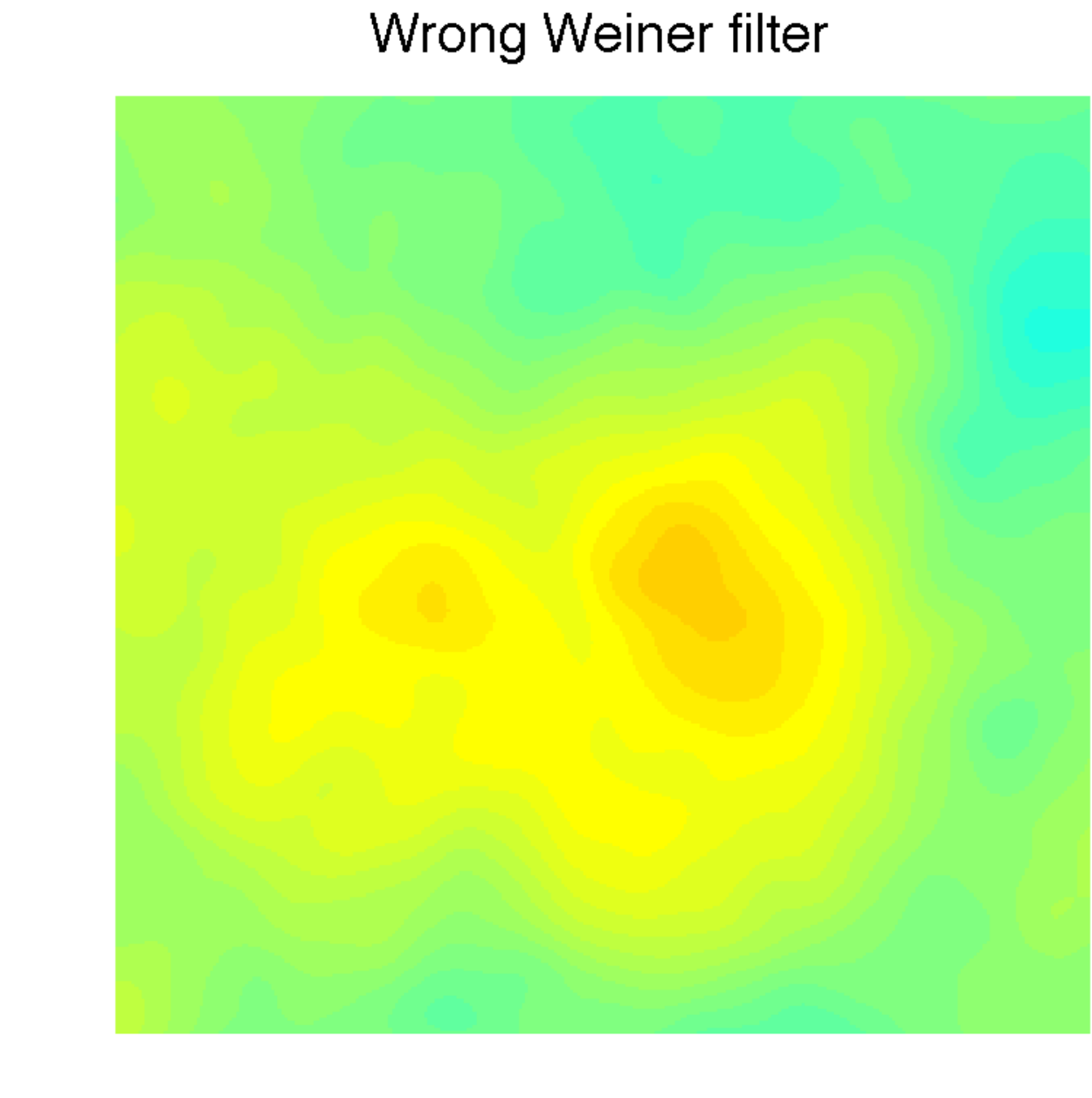}
\includegraphics[height = 1.7in]{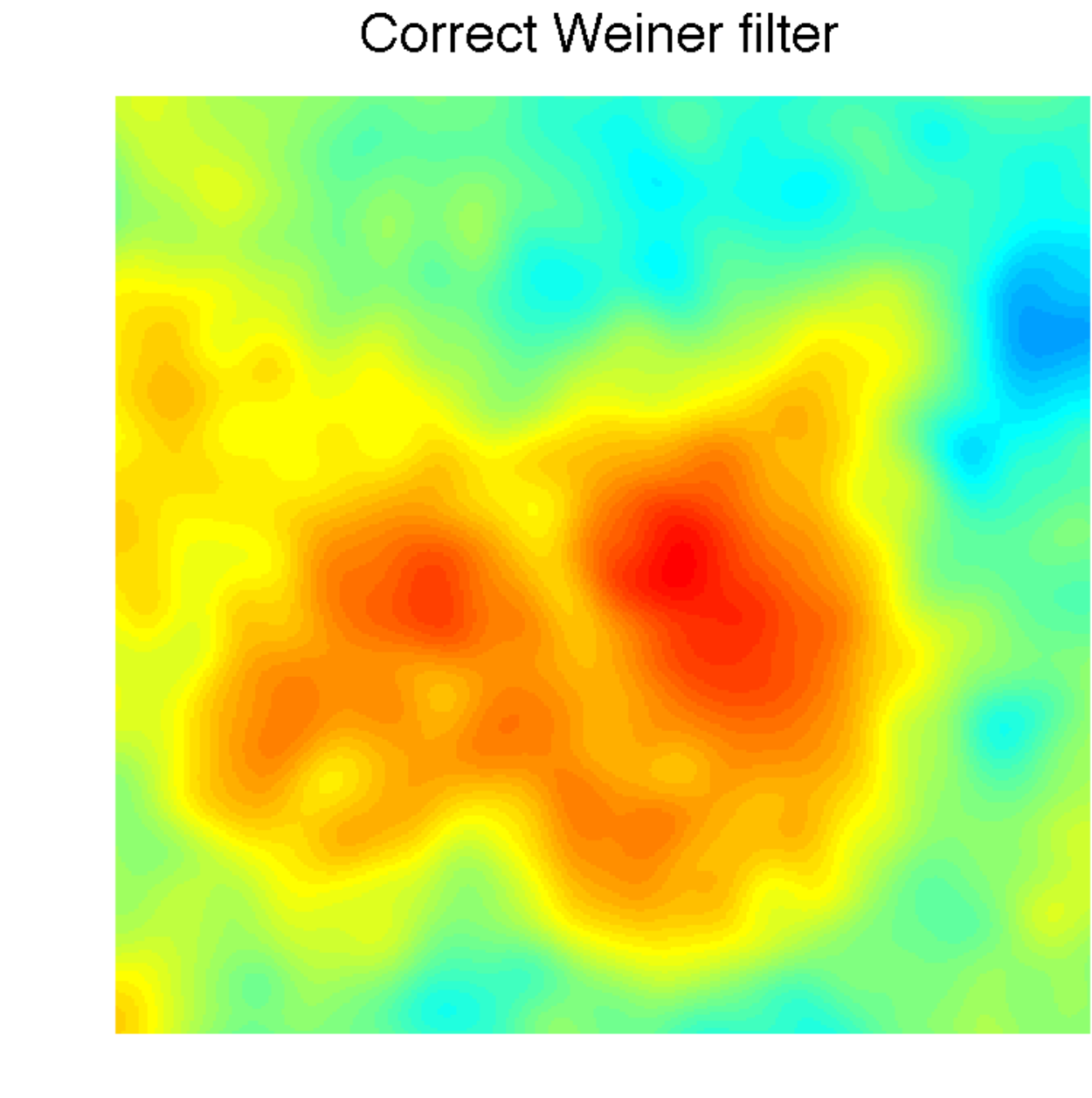}
\includegraphics[height = 1.7in]{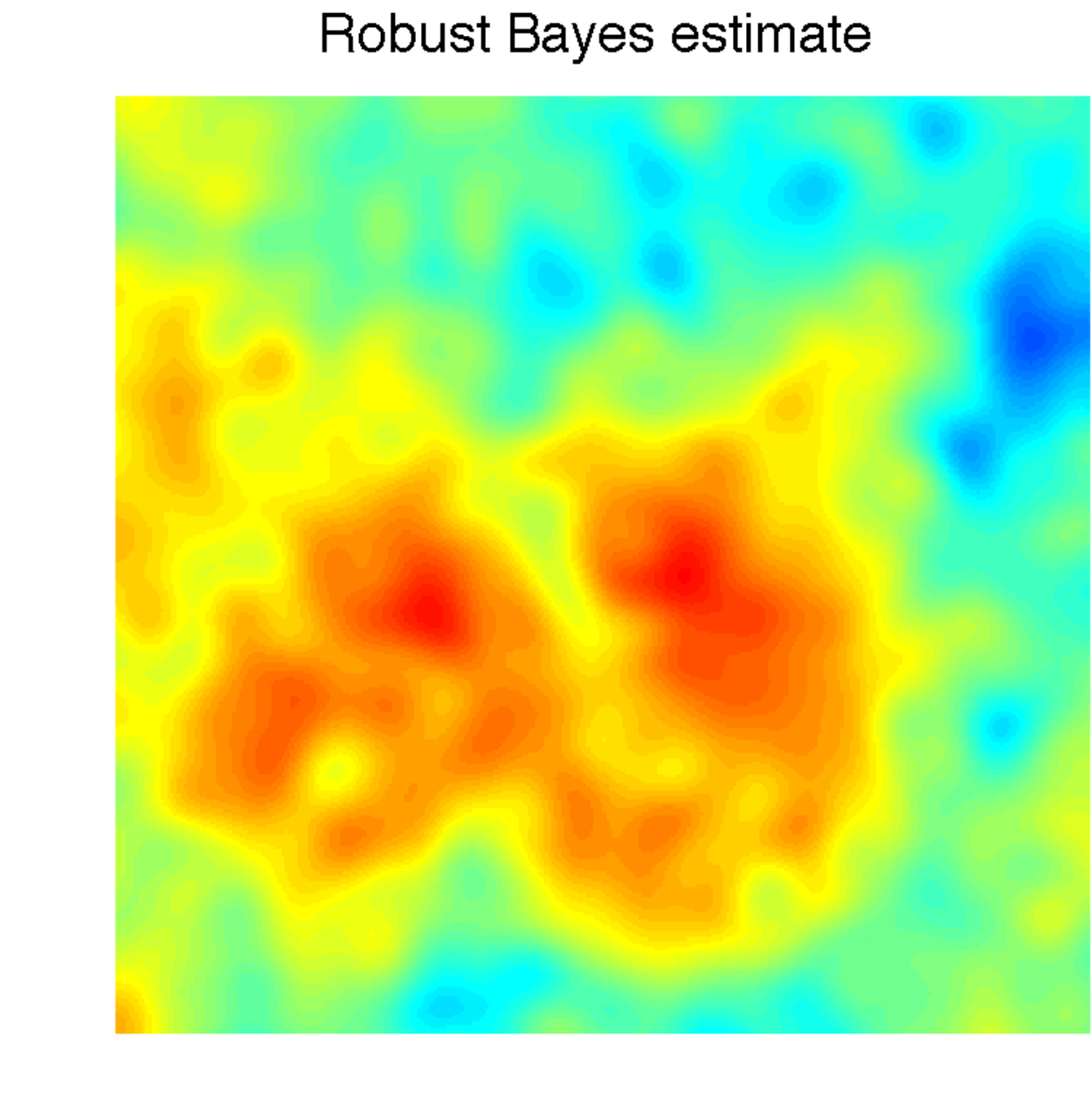}
\caption{{\bf Example 1:}
Pixel space estimates (bottom row) of the filtered lensing potential $|\bs L|\phi(\bs L)\bs 1_{|\bs L|\leq 1000}$ (top). The lensed simulation is done with arcmin pixels, a beam FWHM set to 1 arcmin and additive white noise with a standard deviation of 25 $\mu K$-arcmin. 
The quadratic estimate is shown {bottom left} and the robust Bayes estimate is show {bottom right} using an input  fiducial spectral density $C^{\phi\phi}_{L,\text{fid}}$ set $4$ times smaller than the simulation truth spectral density $C^{\phi\phi}_L$.
The bottom middle two plots show the Wiener filter of the quadratic estimate given by  (\ref{weiner}) based on 
  $C^{\phi\phi}_{L,\text{fid}}$ (middle left) and the true but unknown spectral density $C^{\phi\phi}_L$(middle right).  }
\label{fig1}
\end{figure*}

{\em Remark 3:} 
To connect the Bayes estimator (\ref{bff}) with alternative non-Bayesian estimates of $\sff\phi$, it is  instructive to consider the limit as $n\rightarrow \infty$. The cleanest connection occurs when the fiducial model $C^{\phi\phi}_{L,\text{fid}}$ is set to $0$ (so that $\Lambda_\text{fid}=0$) which postulates no lensing in the observed CMB. 
Making the additional assumption $\Sigma = 2\sigma^2I$
 the estimate (\ref{bff}) simplifies to $\sff{\tilde \phi}= \bigl[1-F^{\sff{\hat\phi}}\bigr]\sff{\hat\phi}$. In Appendix \ref{Der} we show
\begin{align}
\bigl[1-F^{\sff{\hat\phi}}\bigr] =  \left[1- \frac{(2n-2)\sigma^2}{ \sff{\hat\phi}^\dagger\sff{\hat\phi}^* }  \right]^+ + o(1)
\label{SteinsConnect}
\end{align}
where $o(1)\rightarrow 0$ uniformly in $\sff{\hat\phi}$ as $n\rightarrow \infty$ and $x^+\equiv \max\{0,x \}$.
The right hand side of (\ref{SteinsConnect}) is the shrinkage factor used in the famous James-Stein shrinkage estimator \cite{JS}. This follows since the real and imaginary parts of $\sff{\hat\phi}$ are modeled as $2n$  independent Gaussian random variables, each with variance $\sigma^2$.  
In fact, for any fixed number $\epsilon>0$, the supremum of  $|o(1)|$  over the region $\frac{\sff{\hat\phi}^\dagger\sff{\hat\phi}^*}{2n-2} \geq \sigma^2+\epsilon$ converges to zero exponentially fast as $n\rightarrow \infty$. Therefore if one changes $2n-2$ in the numerator of (\ref{SteinsConnect}) to, say $2n$, this exponential convergence fails to hold.
 Even more is true: under the same assumptions on $\Lambda_\text{fid}$ and $\Sigma$,
 the results of \cite{bergerPaper2} show that $\sff{\tilde \phi}$ is a minimax and admissible estimate of $\sff \phi$ with respect to the quadratic loss when $n\geq 2$. 
Similar results for (\ref{bff}) and other Bayes estimates have been derived in the statistical literature (see \cite{lehmann2, bergerBook} for a review of the literature).
 
\section{Simulations}
\label{Sim}

 We  present three simulation examples which give an overview of some of the advantages provided by the robust Bayesian procedure developed in Section \ref{SQE2}.  
 The first example demonstrates robustness and compares with the optimal Wiener filter. The second  compares point-wise error quantification with the quadratic estimator. The final example explores spectral power estimation and demonstrates the robust Bayesian method can generate more accurate detection levels than the quadratic version based on spectral density estimation.


\subsection*{Example 1}

\begin{figure*}[t]
\includegraphics[height = 2.4in]{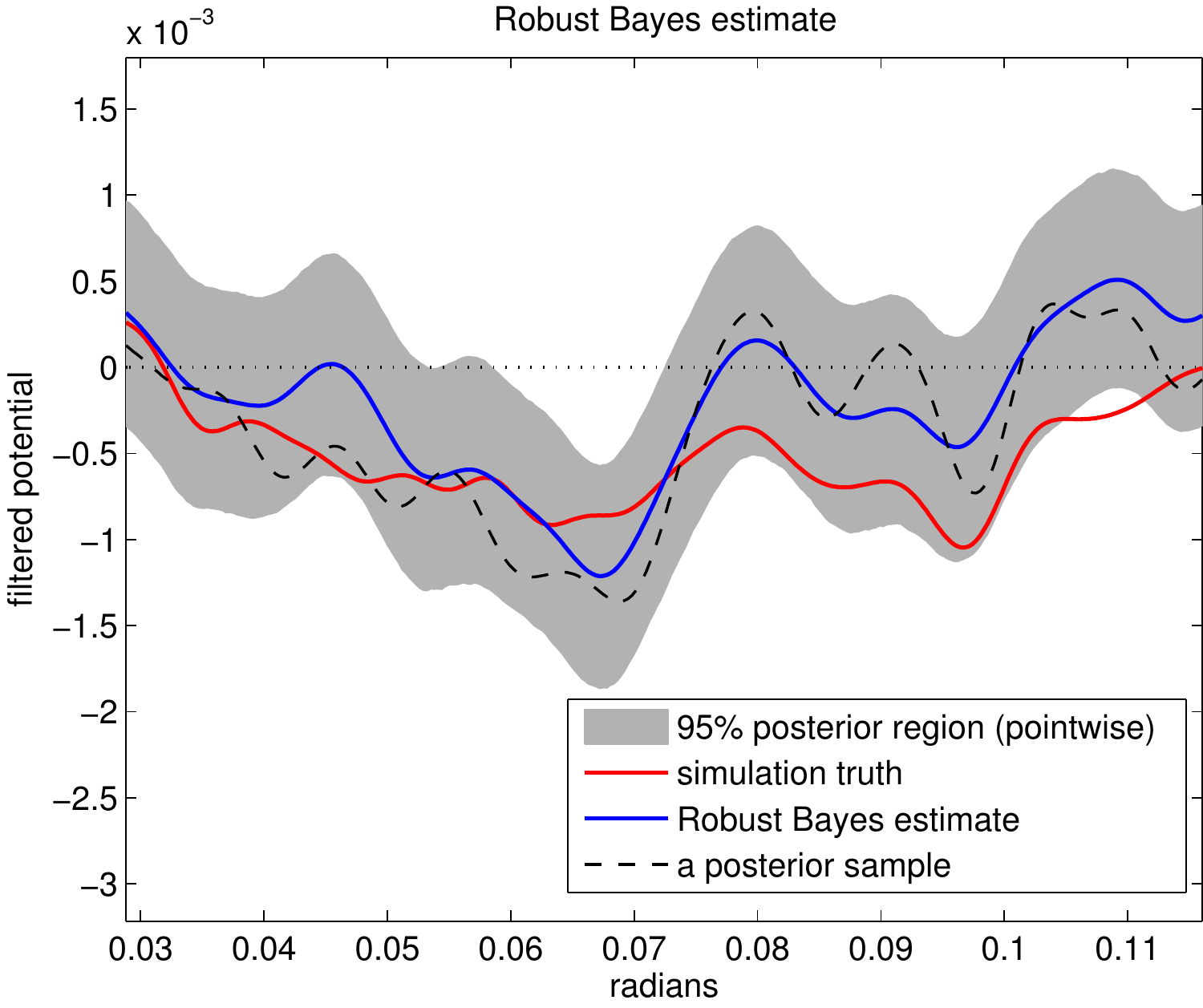}
\includegraphics[height = 2.4in]{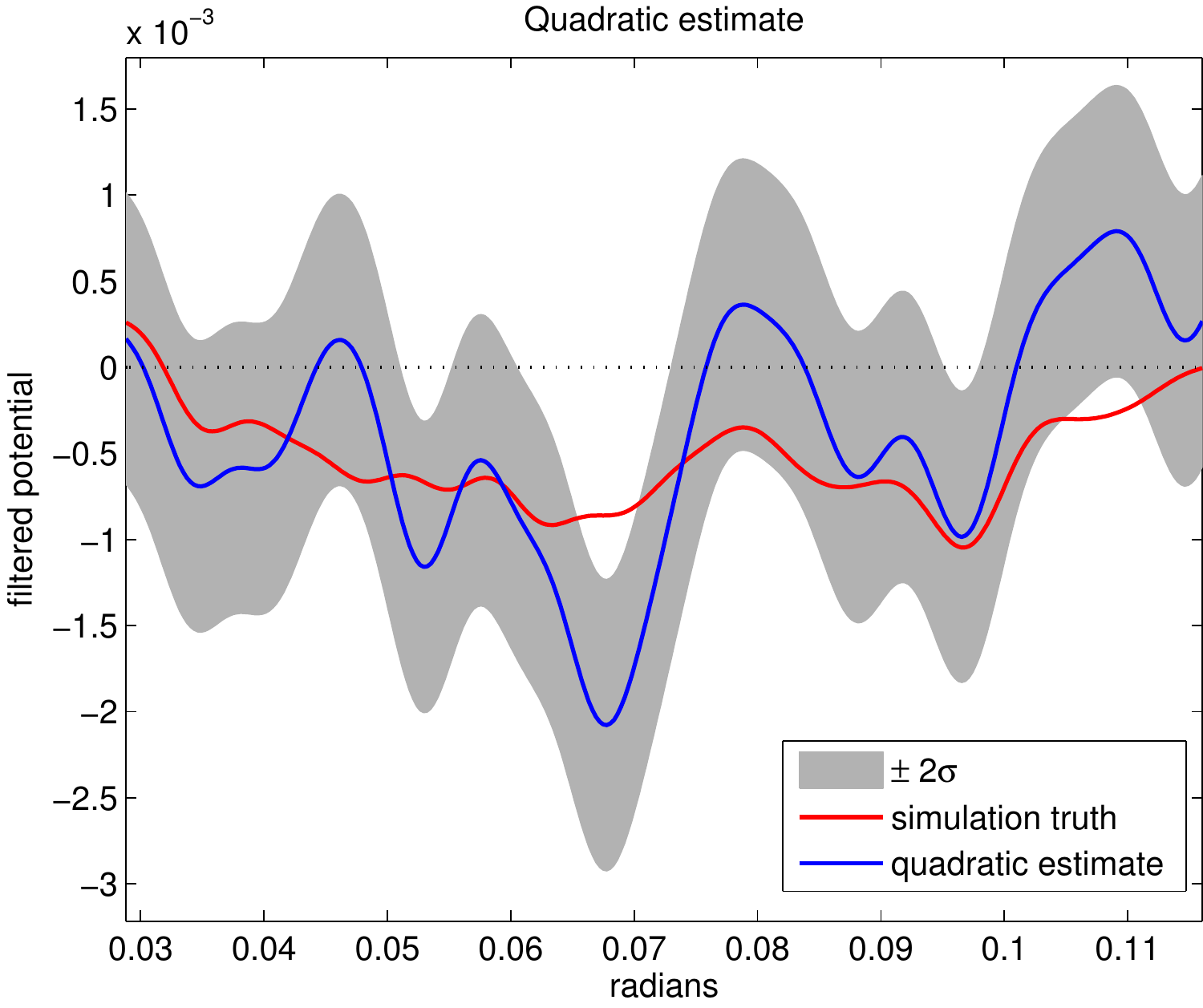}
\caption{{\bf Example 2:}  
 Pixel space line transect plots of the robust Bayes estimate (left, blue) and quadratic estimate (right, blue) of the filtered lensing potential $|\bs L|\phi(\bs L)\bs 1_{|\bs L|\leq 1000}$ plotted in red for both left and right diagrams. 
 The shaded region corresponds to $2\sigma$  error bars on the right and  a point-wise 95\% posterior region on the left.
The lensed simulation is done with arcmin pixels,  a beam FWHM set to 4 arcmin and additive white noise with a standard deviation of 5 $\mu K$-arcmin.  The input  fiducial spectral density $C^{\phi\phi}_{L,\text{fid}}$ used for the robust Bayes estimate is set  $4$ times larger than the simulation true spectral density.}
\label{Slice}
\end{figure*}

In this example we consider the problem of  imaging estimated over-densities  (in pixel space)  of the filtered potential. 
We focus on the filter $|\bs L|\phi(\bs L)\bs 1_{|\bs L|\leq 1000}$ where $\bs 1_{|\bs L|\leq 1000}$ is the top hat indicator with radius $1000$. We remove the spectral power beyond $1000$ for two reasons. First, the high frequency power must be removed from the quadratic estimate or else the noise  completely dominates the image. Secondly the, so called, $N^{(1)}_L$ bias starts to significantly contaminate the results beyond $1000$ (see \cite{Kesden2,Han2010}).

The bottom row of images in FIG.~\ref{fig1} show four different estimates of the filtered lensing potential  in pixel space based on a lensed CMB simulation in a $17^o\times 17^o$ patch of periodic flat sky observed on arcmin pixels. The lensed CMB simulations are done using a Gaussian beam FWHM of 1 arcmin and additive white noise with standard deviation $25\, \mu K$-arcmin (further simulation details can be found in Appendix \ref{SimDets}).   The top plot shows the simulation truth, zoomed in on an over dense region. The bottom left shows the corresponding region for the quadratic estimate and the bottom right shows the robust Bayesian estimate given in (\ref{EQQ}) where the fiducial spectral density $C^{\phi\phi}_{L, \text{fid}}$  is $4$ times smaller than simulation truth spectral density $C^{\phi\phi}_L$. The middle two plots show Wiener filters of the quadratic estimate given by (\ref{weiner}) based on the true, but unknown spectral density $C^{\phi\phi}_L$(middle right) and on the same fiducial spectral density $C^{\phi\phi}_{L, \text{fid}}$ (middle left) used to generate the robust Bayes estimate. 

There are three things to notice here. First,  the robust Bayes estimate successfully shrinks the noisy high frequency terms in the quadratic estimate (i.e.\!\! far left is noisier than than far right). Secondly, if one simply used the  Wiener filter (\ref{weiner}) based on the wrong fiducial model $C^{\phi\phi}_{L, \text{fid}}$ one would seriously over shrink the quadratic estimate (i.e.\! middle left over shrinks). Three, the adaptivity factor $F^{\hat\phi}_L$ is adjusting the robust Bayes estimate to behave more like the optimal Wiener filter (\ref{weiner}) when one has access to the true spectral density 
$C^{\phi\phi}_L$ (i.e.\! middle right and far right look similar). This demonstrates some robustness to mis-specification of the fiducial model.

\subsection*{Example 2}

The second simulation compares point-wise error quantification for estimating the same filtered potential $|\bs L|\phi(\bs L)\bs 1_{|\bs L|\leq 1000}$  as in last example.  To contrast with the last example, we use a different fiducial spectral density: one that is $4$ times too large rather than $4$ times too small. We also change the  beam FWHM to  4 arcmin and additive noise  standard deviation to $5\, \mu K$-arcmin (the pixel size and sky coverage is the same as the last example).
FIG.~\ref{Slice} shows line transect plots of the filtered simulation truth, plotted in red for both left and right diagrams. The right diagram shows the quadratic estimate (blue) with $2\sigma$  error regions shaded grey. The $2\sigma$ region is computed using the Fourier space variance approximation $2|\bs L|^2 A_{\bs L}\bs 1_{|\bs L|\leq 1000}$ for the filtered quadratic estimate. The left diagram shows the robust Bayesian estimate (blue) with the shading region denoting a 95\% the posterior  region (point-wise) which is determined from simulation. We also included a plot of one posterior posterior sample (dashed).  The point-wise error bars corresponding to the Bayes estimate are somewhat smaller than those from the quadratic estimator. However, the main feature of these plots is that, even though the fiducial model is 4 times too large, the Bayes estimate successfully shrinks the noisy high frequency terms in the quadratic estimator. One can also see the advantage of having posterior realizations of possible truths supported by the data (dashed) which allow joint quantification of uncertainty rather than simple point-wise mean and standard error bars provided by the quadratic estimate.

\begin{figure*}[t]
\includegraphics[height = 2.7in]{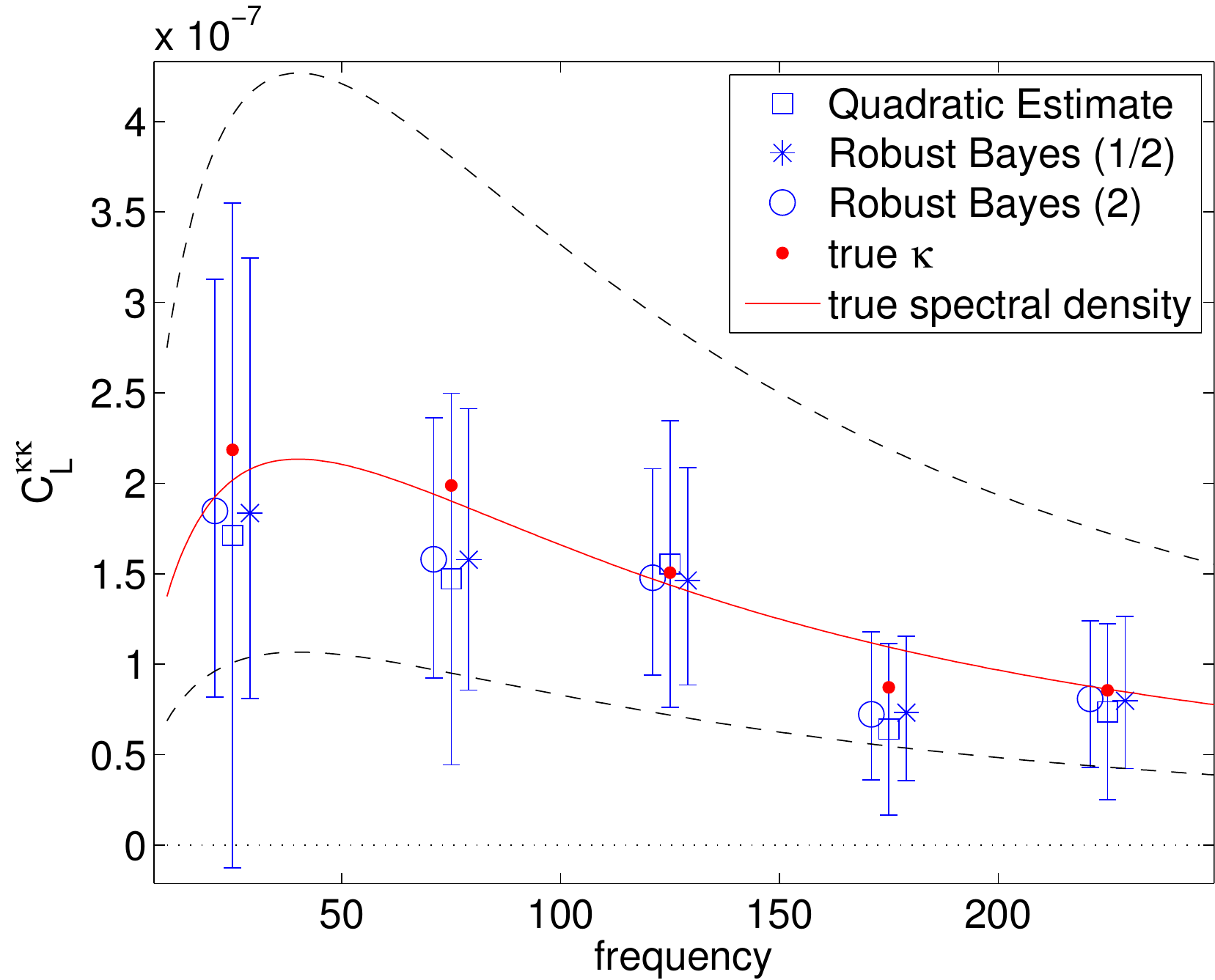}
\includegraphics[height = 2.7in]{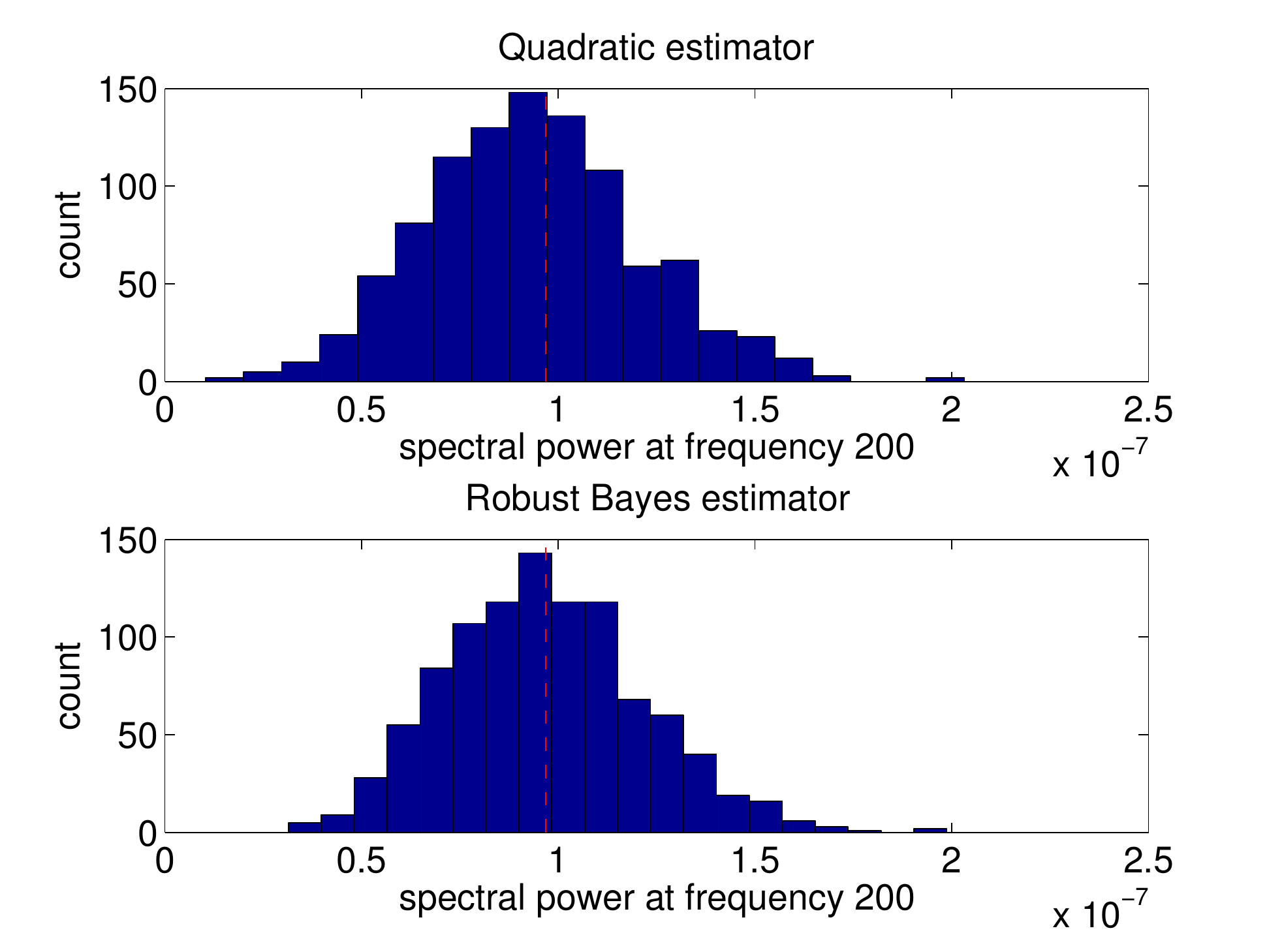}
\caption{{\bf Example 3:} The left plot illustrates spectral power estimation for the convergence $\kappa(\bs L)$ using the quadratic estimate (squares) and the robust Bayesian method (circles and stars) for two different input fiducial models (dashed lines). The error bars for the Bayesian estimates correspond to 2.5 and 97.5 posterior percentiles, whereas the quadratic estimate is displayed with $2\sigma$ error bars  (see Section \ref{Sim} for simulation details). The histograms, at right, show the quadratic estimates and the Bayes estimates of the spectral power in $\kappa(\bs L)$ at wavelength $L=200$ based on 1000 independent lensing simulations. The dashed red vertical lines in the histogram plots designate the value of the true spectral density $C^{\kappa\kappa}_{200}$. See Section \ref{Sim} for further details.
}
\label{aaaa}
\end{figure*}

\subsection*{Example 3}

 In this example we explore spectral power detection and demonstrate that the robust Bayesian method can generate more accurate detection levels than the quadratic version based on spectral density estimation.  CMB gravitational lensing surveys often focus on detecting non-zero spectral density power in the so called convergence field $\kappa(\bs x)\equiv-\nabla^2 \phi(\bs x)/2$ (using the notation found in \cite{ZaldSel1999}) which is a tracer of projected mass. Typically, the quadratic estimate is used to estimate the spectral density $C^{\kappa\kappa}_L$ which, along with standard error bars,  are then used to report $\sigma$-detection levels.
 However, when estimating the spectral density there is always unavoidable cosmic variance, which limits the $\sigma$-detection level achievable. To illustrate this fact, consider  the case that  one is able to observe the  $\kappa(\bs x)$ field directly, with no noise. In this case there is no uncertainty in lensing detection. However,  there is still cosmic variance uncertainty for estimating the spectral density $C^{\kappa\kappa}_L$. We explore this issue by taking  advantage of the robust Bayes method  which easily produces estimates of the spectral power in $\kappa$ directly, instead of through $C^{\kappa\kappa}_L$. We demonstrate better $\sigma$-detection levels for gravitational lensing than would be obtained through quadratic estimates of $C^{\kappa\kappa}_L$.
 
 The parameter of interest, in this case, is the spectral power in $\kappa$. Under the isotropic assumption on $\kappa$ it is natural to radially average the spectral power over concentric annuli about the origin. We only consider annuli which are cut in half since $\kappa(-\bs L)=\kappa(\bs L)^*$, because $\kappa(\bs x)$ is a real random field, which implies redundant information on the other half of the annuli.
 In particular, for each half annuli $\mathcal A$,  we want to estimate the quantity
\begin{equation}
\label{tttt}
\frac{\Delta \bs L}{\# \mathcal A}\sum_{\bs L\in \mathcal A} |\kappa (\bs L)|^2 
\end{equation}
where $\Delta \bs L$ denotes the area of the grid spacing in Fourier space arising from finite sky observations, and $\#\mathcal A$ denotes the number of observed Fourier frequencies in $ \mathcal A$.  Notice the factor $\Delta \bs L$ makes (\ref{tttt}) an unbiased estimate of $C^{\kappa\kappa}_L$ when the  annuli radii and $\Delta \bs L$ are infinitesimally small (this follows by the equality $\langle \kappa(\bs L)\kappa(\bs L^\prime)^*\rangle_{\phi}=\delta_{\bs L - \bs L^\prime} C^{\kappa\kappa}_L$ where $\langle \cdot \rangle_{\phi}$ denotes expectation with respect to $\phi$).

The Bayesian estimate for (\ref{tttt}) is exceedingly easy to construct:  simply average the quantity (\ref{tttt}) computed on each posterior sample  $\phi$ obtained from  Algorithm \ref{alg2}. 
To compare with the quadratic estimate of the spectral density $C^{\kappa\kappa}_L$, which is typically used for detection, we use the following  unbiased (up to leading order) quadratic estimate of $C^{\kappa\kappa}_L$
\begin{equation}
\label{eeest}
\widehat{C^{\kappa\kappa}_L}\equiv\frac{\Delta \bs L}{\# \mathcal A}\sum_{\bs L\in \mathcal A} |\hat\kappa (\bs L)|^2 -\frac{1}{\# \mathcal A}\sum_{\bs L\in \mathcal A}  |\bs L |^4 A_{\bs L}/2
 \end{equation}
 where  $\hat\kappa(\bs x)\equiv-\nabla^2 \hat\phi(\bs x)/2$ and $\hat\phi$ is the quadratic estimate.
Under the Gaussian approximation one gets the following variance for $\widehat{C^{\kappa\kappa}_L}$ 
\begin{align}
\label{vvar} 
\text{var}\,\widehat{C^{\kappa\kappa}_L} &= \frac{1}{(\# \mathcal A)^2}\sum_{\bs L\in \mathcal A} \left( {C^{\kappa\kappa}_{\bs L}} + |\bs L|^4 A_{\bs L}/2 \right)^2 
\end{align}
 This approximation agrees with \cite{Kesden2} and \cite{Han2010} (in the curved sky) when $2A_{\bs L}$ is rotationally symmetric (the missing factor of $2$ found in \cite{Kesden2} and \cite{Han2010} appears since our $\mathcal A$ is only half an annuls). Notice that to compute the variance of $\widehat{C^{\kappa\kappa}_L}$ one must know  the unknown spectral density $C^{\kappa\kappa}_L$. To approximate $\text{var}\,\widehat{C^{\kappa\kappa}_L} $ we therefore replace ${C^{\kappa\kappa}_L}$ in the right hand side of equation (\ref{vvar}) with the estimate $\widehat{C^{\kappa\kappa}_L}$.

Our first comparison of the Bayes and quadratic estimates of (\ref{tttt}) is shown in the left plot of FIG.~\ref{aaaa} and is based on one realization of a lensed CMB field ($17^o\times 17^o$ periodic sky, arcmin pixels,  beam FWHM = 1 arcmin, white $25\, \mu K$-arcmin noise) with spectral density $C^{\kappa\kappa}_L$  shown in red. 
We computed two Bayes estimates: one which uses a fiducial model $C^{\phi\phi}_{L,\text{fid}}$ which is 2 times too large (estimates are denoted with circles, fiducial model with dashed line), and one using $C^{\phi\phi}_{L,\text{fid}}$ which is 2 times too large (estimates are denoted with stars, fiducial model with dashed line). The estimates are based on the first $5$ concentric half annuli (width = 50) about the origin.   The attached error bars correspond  to $2.5^\text{th}$ and $97.5^\text{th}$ posterior percentiles.  
The squares show  the quadratic estimate  (\ref{eeest}) with $2\sigma$ error bars based on  (\ref{vvar}).
There are two points here. First, the  Bayesian error bars are smaller and more naturally handle the positivity constraint. Secondly, the Bayesian regions are relatively robust with respect to misspecification of $C^{\phi\phi}_{L,\text{fid}}$ (i.e.\! the error bars attached to the circles and the stars are about the same size).

For our second comparison  we  check the posterior coverage probabilities, the quadratic estimate coverage probabilities and the relative sizes of the quadratic and Bayes error bars. We
simulated 1000 independent lensed CMB realizations (with the same simulation configuration as above) using independent realizations of the unlensed CMB, the noise and the lensing potential for each simulation. The right hand plot of FIG.~\ref{aaaa}  shows the histograms of the corresponding quadratic and Bayes estimates of (\ref{tttt}) at frequency $|\bs L|=200$ (with half annuli width set to $50$). The Bayes estimates were generated using a fiducial spectral density which is 2 times smaller than the input spectral density used for simulation. The dashed red vertical lines designate the value of the true spectral density $C^{\kappa\kappa}_{200}$. 

The two histograms in FIG.~\ref{aaaa} are similar with a slight variance and bias reduction in the Bayes estimates ($15\%$ smaller variance and  $50\%$ smaller bias).   However, the main difference can be found in the associated error bars. For the Bayes estimate we use a posterior region based on $2.5^\text{th}$ and $97.5^\text{th}$ posterior percentiles for estimating (\ref{tttt}), not $C^{\kappa\kappa}_{200}$. Surprisingly, exactly $950$ of the $1000$  posterior regions contained the true value of (\ref{tttt}), which is different in each simulation. Conversely, $934$ out of the $1000$  $2\sigma$-error bars associated with the quadratic estimate covered the true value of $C^{\kappa\kappa}_{200}$. One would expect typical Monte Carlo fluctuations  of about $\pm 7$ (if true coverage was $95\%$) so this gives moderate evidence that the quadratic estimate error bars slightly undercover the truth. Moreover, the  width of the posterior regions are $15\%$ smaller than the corresponding error bars for the quadratic estimate. This is due to the fact that the Bayes estimate does not contain cosmic variance uncertainty between (\ref{tttt}) and $C^{\kappa\kappa}_{200}$.

It should be noted that it is possible to adjust the variance calculation (\ref{vvar}) to represent  the uncertainty for estimating (\ref{tttt}), rather than the uncertainty for $C^{\kappa\kappa}_L$. This is somewhat beside the point, however, that the Bayesian estimates do this naturally no matter what functional one is estimating.  There still needs to be a verification process that the Bayesian method behaves appropriately, however,  for any rigorous scientific application. This most likely would include some type of simulation study. It may also include, as is done in  Remark 3 of Section \ref{CCC}, a derivation of the frequentest properties of the Bayesian estimates.

%
%

\section{Discussion}
\label{Dis}
The unbiasedness constraint in the quadratic estimator forces large variability in the estimated gravitational potential, especially at high frequency. This can present difficulties for exploratory data analysis, mapping the lensing potential and estimating nonlinear functionals of the gravitational potential, for example. In this paper we study the potential advantages obtained  by relaxing the unbiasedness constraint in the quadratic estimator.  To accomplish this we derived a regression characterization of the quadratic estimate which then allows one to clearly see how bias can be introduced for a reduction of variance: through ridge regression and Bayesian techniques. The Bayesian framework is especially cogent, essentially treating the quadratic estimate as data while incorporating prior information on $C^{\phi\phi}_L$. The resulting estimate is an adaptive Wiener filter adjustment to the raw quadratic estimate---shrinking frequencies with small SNR and retaining those with high SNR. As an alternative to a full Bayesian analysis, which can be somewhat demanding, we present a non-informative prior which not only leads to estimates with desirable frequentist properties (such as an asymptotic James-Stein shrinkage behavior) but also yields a posterior distribution which is easy to simulate without resorting to Monte Carlo techniques. 
 The non-informative prior requires the user to input a fiducial model, but is designed to be robust to misspecification of this input model. 

One clear advantage of the Bayesian analysis is the availability of posterior samples to construct estimates and uncertainty quantification for any non-linear function of the gravitational potential, including spectral density estimation. 
Indeed, rather than using the shrinkage  formula (\ref{EQQ}) directly (it becomes numerically unstable) we recommend averaging samples from the posterior obtained from algorithms \ref{alg1} and \ref{alg2} given in Section \ref{CCC}. 
In Section \ref{Sim}  we explored the advantages gained by having easily obtained posterior samples: estimating and quantifying uncertainty in functionals of the gravitational potential and joint quantification of uncertainty. 

A drawback of the above Bayesian analysis---indeed of the quadratic estimate itself---is the bias obtained from the Taylor truncation on the lensed CMB used to derive the quadratic estimate. The spectral properties of this bias is relatively well understood \cite{Kesden2, Han2010} when marginalizing over the randomness inherent in the large scale structure. However, it is unclear how this bias effects other features of the estimate of $\phi$.
 Therefore, when applying the Bayesian methods described in the paper, we recommend avoiding the frequencies which are shown to be contaminated by biases. An alternative to avoiding frequencies is adjust $2A_{\bs L}$ to include the so called $N^{(1)}_{\bs L}$ and $N^{(2)}_{\bs L}$ biases. However, until we have a good understanding of the sensitivity of $N^{(1)}_{\bs L}$ and $N^{(2)}_{\bs L}$ to a fiducial model this remains an unexplored possibility.

\appendix
\section{Simulation Details}
\label{SimDets}
The fiducial cosmology used in our simulations is based on a flat,
power law $\Lambda$CDM cosmological model, with baryon density
$\Omega_b=0.044$; cold dark matter density $\Omega_\text{cdm}=0.21$;
cosmological constant density $\Omega_\Lambda=0.74$; Hubble parameter
$h=0.71$ in units of 100$\,$km$\,$s$^{-1}\,$Mpc$^{-1}$; primordial
scalar fluctuation amplitude $A_s(k=0.002\,$Mpc$^{-1}) = 2.45\times
10^{-9}$; scalar spectral index $n_s(k=0.002\,$Mpc$^{-1}) = 0.96$;
primordial helium abundance $Y_P=0.24$; and reionization optical depth
$\tau_r=0.088$. The CAMB code is used to generate the theoretical
power spectra \cite{CAMB}.

To construct the lensed CMB simulation used in this paper we first  generate a high resolution simulation of $\Theta(\bs x)$ and the gravitational potential $\phi(\bs x)$ on a periodic $17^\text{o} \times 17^\text{o}$ patch of the flat sky with $0.25$ arcmin pixels. The lensing operation is performed by taking the numerical gradient of $\phi$, then using linear interpolation to obtain the lensed field $\Theta(\bs x + \nabla \phi(\bs x))$.  
We down-sample the lensed field, every $4^\text{th}$ pixel to obtain the desired arcmin pixel resolution for the simulation output.  
A Gaussian beam is then applied in Fourier space using FFT of the lensed fields. Finally white noise is added in pixel space.

\section{Derivation of equation (\ref{SteinsConnect})}
\label{Der}

Start by letting $s\equiv {\| \sff{\hat\phi}\|^2_\text{fid} }/(n-1)$  which simplifies to  $\frac{\sff{\hat\phi}^\dagger \sff{\hat\phi}^*}{\sigma^2(2n - 2)}$  when  $\Lambda_\text{fid}=0$ and $\Sigma=2\sigma^2 I$. Notice that `s' behaves like a ratio between the observed signal power and the nominal noise level. 
To make the following calculations more readable we also let $m = n-1$. To establish (\ref{SteinsConnect}) it will be sufficient to show
\begin{equation}
  1 -  \frac{1}{s} \frac{P\left( m+1, ms\right)}{P\left( m, ms \right)} = \left[ 1- \frac{1}{s} \right]^+ + o(1)
  \end{equation}
where $o(1)$ is a function of both $m$ and $s$ such that
\begin{equation}
\label{showme7}
\sup_{s>0} |o(1)| \rightarrow 0 \quad \text{as $m\rightarrow \infty$}.
\end{equation}

Notice that $ 1 -  \frac{1}{s} \frac{P\left( m+1, ms\right)}{P\left( m, ms \right)}$ is increasing in $s$. This was shown in Lemma 2.1.1(vii) of \cite{bergerPaper} where it is noted that the density for the random variable $\xi^{-1}$ has a decreasing monotone likelihood ratio in $s$. This is sufficient for stochastic domination (see Lemma 3.4.2 in \cite{lehmann}, for example) which implies the expected value $\langle \xi^{-1}|\sff{\hat \phi} \rangle \equiv \int \xi^{-1}\mathcal P(\xi|\sff{\hat\phi})=\frac{1}{s} \frac{P\left( m+1, ms\right)}{P\left( m, ms \right)}$ goes down when $s$ goes up. 
Therefore  the supremum $\sup_{0<s\leq 1} |o(1)|$ is attained at $s=1$.  This follows by the fact that $ 1 -  \frac{1}{s} \frac{P\left( m+1, ms\right)}{P\left( m, ms \right)}$  is increasing and  positive (because  it equals $1- \langle \xi^{-1}|\sff{\hat \phi}\rangle$ and $\xi^{-1}$ is supported in $(0,1]$).
Therefore to prove (\ref{showme7}) it will be sufficient to establish $\sup_{s\geq 1} |o(1)|\rightarrow 0$  since
\[ \sup_{0<s\leq 1} |o(1)|\leq  \sup_{s\geq 1} |o(1)|. \]
To study the case $s\geq 1$, integration by parts gives
\begin{align}
1 -  \frac{1}{s} \frac{P\left( m+1, ms\right)}{P\left( m, ms \right)}&=1 -  \frac{1}{s} \frac{P\bigl( m,  m  s\bigr) - \frac{(m s)^{m} e^{- m s}}{m!}}{P\bigl( m,  m  s\bigr)}\nonumber\\
&=1 -  \frac{1}{s}+ \frac{m^m}{m!}\frac{  (s e^{-s})^{m} }{s P\bigl( m,  m  s\bigr)}.\nonumber 
\end{align}
Therefore
\begin{align}
  \sup_{s\geq 1} |o(1)| &= \frac{(m/e)^m}{m!}  \sup_{s\geq1}  \frac{  (s e^{1-s})^{m} }{s P\bigl( m,  m  s\bigr)}\nonumber\\
 &\leq  \frac{(m/e)^m}{m!P\bigl( m,  m \bigr)} \label{convv1}\\
 & \sim \frac{2}{\sqrt{2 \pi m}}\rightarrow 0 \label{convv2}
  \end{align}
Line (\ref{convv1}) follows since  $s e^{1-s} \leq 1$ (with equality only when $s=1$) and the fact that $P(m, x)$ is an integral of a positive function over the interval $(0,x]$. Line (\ref{convv2}) follows  by Stirling's approximation  and the fact that  $P(m,m) = Pr[\sum_{k=1}^m X_k \leq m]$ where $X_k$ are independent Gamma random variables with mean and variance $1$.
The central limit theorem then says $P(m,m)=Pr[\sqrt{m}\sum_{k=1}^m (X_k - 1) \leq 0]\rightarrow Pr [Z\leq 0]=1/2$ where $Z$ is a standard normal random variable.

\section{FFT and the quadratic estimator}
\label{programing}

Unfortunately the regression format is not amenable to computation. Therefore one must take advantage of the Fourier filtering characterization of the quadratic estimator to implement it on the computer. There are a few details that we mention here. The nominal form of the quadratic estimator is 
\[\hat \phi(\bs L)=A_{\bs L} \int_{\Bbb R^d}    \tilde \Theta(\bs \ell+\bs L)  \tilde \Theta(\bs \ell)^*g_{\bs L}(\bl)\, {d\bl}\]
 where the weights $g$ must satisfy the constraint $g_{\bs L}(\bl) f_{\bs L}(\bl)\geq$ and the normalizing constant $A_{\bs L}^{-1}\equiv\int  {g_{\bs L}(\bl) f_{\bs L}(\bl)} \,d\bl $ ensure unbiasedness. In \cite{Hu2001b} the optimal weights $g$ are derived to  be proportional to ${f_{\bs L}(\bl)^*}/[{C^{\Theta\Theta}_{\bl+\bs L,\text{expt}}C^{\Theta\Theta}_{\bl,\text{expt}}}] $. However there may be cases where some pairs of frequencies $(\bs \ell, \bs \ell+\bs L)$ need to be excluded. For example, when $\tilde \Theta$ is approximated using a discrete FFT one wants to avoid using the frequencies that have a large amount of aliasing:  when $|\ell_1|>2\pi/ (2 \Delta x_1)$ or  $|\ell_2|>2\pi/ (2 \Delta x_2)$  (the Nyquist limits, where $\Delta x_1$ is the grid spacing in the first coordinate  position space).  In addition we want to avoid using $\tilde \Theta(0)$ (since it contains no information on lensing). To handle this we introduce a  masking function $M(\bl)$ and absorb it into the optimal $g$
 \[ g_{\bs L}(\bl)\equiv \frac{f_{\bs L}(\bl)^*}{C^{\Theta\Theta}_{\bl+\bs L,\text{expt}}C^{\Theta\Theta}_{\bl,\text{expt}}} M(\bl)M(\bl+\bs L). \]
 In our case we use the following mask
 \[M(\bl)=\bs 1_{|\ell_1|< \frac{\pi}{ 2 \Delta x_1}} \bs 1_{|\ell_2|< \frac{\pi}{ 2 \Delta x_2}} \bs 1_{\bl \neq 0}\]
to avoid aliasing errors  (we cut at half the Nyquist) and the $0$ frequency.  This masking slightly alters the gradient filtering characterization of the quadratic estimate found in \cite{Hu2001b}. Under the assumption that the masking function satisfies  $M(\bl)=M(-\bl)$ one can derive following characterization of the quadratic estimator:
\begin{align}
\label{FFTTT}
 \hat\phi(\bs L)&= -2 {i A_{\bs L}\,\bs L} \cdot \int  e^{-i\bs x\cdot \bs L} \vec G(\bs x)W(\bs x) \frac{d\bs x}{2\pi}
  \end{align}
where $\vec G(\bl)\equiv  \frac{i \bl  C_{\bl }^{\Theta\Theta}  \varphi(\bl )^* M(\bl)  }{C^{\Theta\Theta}_{\bl,\text{expt}} } \tilde \Theta( \bl) $ and $W(\bl)\equiv \frac{ \varphi(\bl )^*M(\bl)}{ C^{\Theta\Theta}_{\bl,\text{expt}} } \tilde \Theta(\bl)$.

Since the masking function is not radially symmetric, the normalization factor $A_{\bs L}$ will not be either. Since a nominal Riemann approximation to the integral is computational intensive (a two dimensional integral is required for each $\bs L$) we derive a Fourier representation similar to (\ref{FFTTT}) which can utilize a Fast Fourier Transform. 
The normalizing factor becomes
\begin{align*}  
A_{\bs L}^{-1}&\equiv\int  {g_{\bs L}(\bl) f_{\bs L}(\bl)} \,d\bl \\
&=\int  \frac{|f_{\bs L}(\bl)|^2}{C^{\Theta\Theta}_{\bl+\bs L,\text{expt}}C^{\Theta\Theta}_{\bl,\text{expt}}}  M(\bl)M(\bl+\bs L) \,d\bl \\
&=\sum_{k,j=1}^d \frac{ L_k L_j}{\pi}  \int e^{-i\bs x\cdot\bs L} [A(\bs x) C_{k,j}(\bs x) - B_{k}(\bs x) B_j(\bs x)]  \frac{d\bs x}{2\pi} 
\end{align*}
where $A(\bl)\equiv \frac{|\varphi(\bl)|^2 M(\bl)}{C^{\Theta\Theta }_{\bl, \text{expt}} }$, $B_k(\bl)\equiv  i\ell_k C^{\Theta\Theta}_\bl A(\bl)$ and $C_{k,j}(\bl)\equiv   \ell_j\ell_k (C^{\Theta\Theta}_\bl)^2 A(\bl)$
for $k,j=1,2$. 

%
%
%

%
%

\begin{acknowledgments}
We gratefully acknowledge enlightening discussion with Jim Berger, Lloyd Knox and Alexander van Engelen.
\end{acknowledgments}


\end{document}